\documentclass[twocolumn,tighten]{aastex631}

\usepackage{amsmath}
\usepackage{natbib}
\usepackage{xspace}
\usepackage{mhchem}

\newcommand{\code}[1]{\texttt{#1}\xspace}
\newcommand{\Gaia}{\textit{Gaia}}
\newcommand{\starfit}{\texttt{StarFit}}
\newcommand{\degree}{$^{\circ}$}
\newcommand{\kmsec}{\mbox{km~s$^{\rm -1}$}}

\newcommand{\msun}{\mbox{$M_{\odot}$}}

\newcommand{\RN}[1]{%
  \textup{\uppercase\expandafter{\romannumeral#1}}%
}
\newcommand{\rpro}{\mbox{{\it r}-process}\xspace}

\newcommand{\spro}{\mbox{{\it s}-process}\xspace}

\newcommand{\teff}{\mbox{$T_{\rm eff}$}}
\newcommand{\logg}{\mbox{log~{\it g}}}
\newcommand{\vt}{\mbox{$v_{\rm t}$}}

\newcommand{\feh}{\mbox{[Fe/H]}}

\newcommand{\LCDM}{\mbox{$\Lambda$CDM}}

\newcommand{\MITPhysics}{MIT Department of Physics, 
77 Massachusetts Avenue, Cambridge, MA 02139, USA}
\newcommand{\MITKavli}{MIT Kavli Institute for Astrophysics and Space Research, 
77 Massachusetts Avenue, Cambridge, MA 02139, USA}
\newcommand{\UChicagoPhysics}{Department of Astronomy \& Astrophysics, University of Chicago, 5640 S. Ellis Avenue, Chicago, IL 60637, USA}
\newcommand{\UChicagoKavli}{Kavli Institute for Cosmological Physics, University of Chicago, Chicago, IL 60637, USA}

\newcommand*{\noFillCircle}{\ensuremath{\circ}} % Empty circle
\newcommand*{\fullFillCircle}{\ensuremath{\bullet}} % Full circle

\begin{document}

%\title{Chemical Abundances of Very Metal-Poor Stars in the core and stream\\ of the Sagittarius Dwarf Galaxy}
%\title{The enrichment histories of core and stream of the Sagittarius Dwarf Galaxy are ????}
\title{Early $r$-process Enrichment and Hierarchical Assembly Across the Sagittarius Dwarf Galaxy\footnote{This paper includes data gathered with the 6.5 meter Magellan Telescopes located at Las Campanas Observatory, Chile.}}

\shorttitle{$r$-process Enrichment Across the Sagittarius Dwarf Galaxy}
\shortauthors{Ou et al.}

\author[0000-0002-4669-9967]{Xiaowei Ou}
\affiliation{\MITPhysics}
\affiliation{\MITKavli}

\author[0000-0002-1462-0265]{Alexander Yelland}
\affiliation{\MITPhysics}
\affiliation{\MITKavli}

\author[0000-0002-7155-679X]{Anirudh Chiti}
\affiliation{\UChicagoPhysics}
\affiliation{\UChicagoKavli}

\author[0000-0002-2139-7145]{Anna Frebel}
\affiliation{\MITPhysics}
\affiliation{\MITKavli}

% \author[0000-0002-4863-8842]{Alex Ji}
% \affiliation{\UChicagoPhysics}
% \affiliation{\UChicagoKavli}

\author[0000-0002-9269-8287]{Guilherme Limberg}
\affiliation{\UChicagoPhysics}
\affiliation{\UChicagoKavli}

\author[0000-0001-9178-3992]{Mohammad K.\ Mardini}
\affiliation{\MITKavli}
\affiliation{Department of Physics, Zarqa University, Zarqa 13110, Jordan}
\affiliation{Jordanian Astronomical Virtual Observatory, Zarqa University, Zarqa 13110, Jordan}
\affiliation{Joint Institute for Nuclear Astrophysics -- Center for the Evolution of the Elements (JINA-CEE), USA}

\begin{abstract}

Dwarf galaxies like Sagittarius (Sgr) provide a unique window into the early stages of galactic chemical evolution, particularly through their metal-poor stars. 
By studying the chemical abundances of stars in the Sgr core and tidal streams, we can gain insights into the assembly history of this galaxy and its early heavy element nucleosynthesis processes.
We efficiently selected extremely metal-poor candidates in the core and streams for high-resolution spectroscopic analysis using metallicity-sensitive photometry from SkyMapper DR2, and \Gaia\ DR3 XP spectra and proper motions.
% (Section~\ref{sec:obs})
% This allowed us to obtain a high-purity selection of Sgr members based on stellar kinematics while reducing the chances of potential contamination from the Milky Way halo.
We present a sample of 37 Sgr stars with detailed chemical abundances, of which we identify 10 extremely metal-poor (EMP; $\ce{[Fe/H]} \le -3.0$) stars, 25 very metal-poor (VMP; $\ce{[Fe/H]} \le -2.0$) stars, and 2 metal-poor (MP; $\ce{[Fe/H]} \le -1.0$) stars.
This sample increases the number of extremely metal-poor Sgr stars analyzed with high-resolution spectroscopy by a factor of five.
Of these stars, 15 are identified as members of the Sgr tidal stream, while the remaining 22 are associated with the core.
We derive abundances for up to 20 elements and identify no statistically significant differences between the element abundance patterns across the core and stream samples.
% In particular, we examined the origin of two stream stars with significantly lower $\alpha$ elements and three \rpro\ enhanced stars in detail, evaluating if they were form in-situ or accreted in Sgr (Section~\ref{sec:discussion}).
Intriguingly, we identify stars that may have formed in ultra-faint dwarf galaxies that accreted onto Sgr, in addition to patterns of C and \rpro\ elements distinct from the Milky Way halo.
% also found in other Milky Way dwarf systems
Over half of the sample shows a neutron-capture element abundance pattern consistent with the scaled solar pure \rpro\ pattern, indicating early \rpro\ enrichment in the Sgr progenitor.
% (Section~\ref{sec:summary})

\end{abstract}



%%%%%%%%%%%%%%%%%%%%%%%%%%%%%% BODY OF PAPER %%%%%%%%%%%%%%%%%%%%%%%%%%%%%%%%%%%

%%%%%%%%%%%%%%%%%%%%%%%%%%%%%%%%%%%%%%%%%%%%%%%%%%
\section{Introduction} \label{sec:intro}

% INTRODUCTION TO CHEMICAL EVOLUTION AND METAL-POOR STARS IN DWARF GALAXIES
Understanding the chemical evolution of galaxies is fundamental for analyzing how elements are synthesized and distributed over cosmic timescales \citep[see e.g.,][]{burbidge57,timmes95,beers05,nomoto13,maiolino19,kobayashi20}.
Through cycles of star formation and stellar nucleosynthesis in a galaxy, its stars, supernovae, and compact remnants enrich the interstellar medium with heavier elements \citep[see e.g.,][]{chiappini97,tacconi20}.
Each new generation of stars forms with the chemical composition of their surrounding gas.
Therefore, ancient, metal-poor stars serve as unique probes of the early stages of galactic chemical enrichment \citep[see e.g.,][]{beers05,frebel15}.
By examining the chemical compositions of metal-poor stars in the Milky Way's satellite dwarf galaxies, we can learn about the evolution, assembly history, and early chemical enrichment of small-scale systems \citep[see e.g.,][]{tolstoy09,kirby13,helmi20,deason24}.

Dwarf galaxies exhibit a wide range of physical properties, including total stellar mass, star formation histories, and environmental interactions \citep{mcconnachie12,pace24}.
Consequently, the chemical evolution of these systems is strongly influenced by the varying rates of star formation, metal mixing, and the frequency of enrichment events such as supernovae, neutron star mergers, or asymptotic giant branch (AGB) star contributions \citep{tolstoy09}. 
% OTHER PEOPLES WORK AND LIMITATIONS
Recent studies have begun to focus on the evolution of neutron-capture elements as they can provide specific details about the formation history and early star formation \citep{cohen09,skuladottir19,skuladottir20,hirai24}. 
In particular, europium abundances in low-metallicity stars were measured in disrupted dwarf systems \citep{matsuno21,naidu22,ernandes24,ou24c,limberg24,atzberger24} and globular clusters \citep{monty24,ceccarelli24}.
Several of these measurements exhibited a clear rise in [Eu/Fe] with increasing metallicity; a trend that has been attributed to a delayed \rpro\ source, such as a neutron star merger, as the origin of europium \citep{naidu22,frebel23,limberg24,ernandes24,ou24c}.
% For the Gaia-Sausage/Enceladus (GSE) dwarf galaxy \citep{}, as well as several other intact/disrupted dwarf systems \citep{matsuno21,naidu22,ernandes24,ou24c,limberg24,chiti24} and globular clusters \citep{monty24,ceccarelli24}, europium abundances in low-metallicity stars were measured and exhibited a clear rise in [Eu/Fe] with increasing metallicity, up to $\rm{[Fe/H]}\sim-1.0$.
% This trend, and the origin of europium, has been attributed to a delayed \rpro\ source, such as a neutron star merger.
Similarly, the ultra-faint dwarf (UFD) Reticulum\,II shows strong \rpro\ enhancement in the majority of its stars, also suggesting the operation of early neutron star mergers \citep[see e.g.,][]{ji16}.
Nevertheless, the overall quantity of available europium measurements remains sparse in dwarf galaxies.

% SAGITTARIUS PROPERTIES
In this study, we focus on the Sagittarius (Sgr) dwarf galaxy and its associated stellar stream.
Sgr is located near the inner Galactic region of the Milky Way and is currently disrupting \citep{ibata94, ibata95}.
The core of Sgr is located approximately $\sim 25 \; \rm{kpc}$ away \citep{ibata97,mateo98,ferguson20}.
However, over the past $\sim 5 \; {\rm Gyr}$, Sgr has undergone significant tidal stripping due to its interactions with the Milky Way, resulting in the formation of two prominent stellar streams \citep{majewski03}, winding twice around the Milky Way and extending across the sky.
Though it remains one of the Milky Way's most massive dwarf galaxies, this ongoing tidal stripping of Sgr has torn a significant amount of stellar mass away from the core.
Today, the stellar mass of Sgr is $\sim 10^{8} \, M_{\odot}$ \citep{vasiliev20}, whereas prior to tidal disruption, it is estimated to have been up to $\sim 10^{10} \, M_{\odot}$ \citep{niederste-ostholt10}.
Hence, Sgr provides a unique opportunity to investigate the erstwhile internal radial distribution of chemical enrichment process(es) in the early Sgr system and to use its present-day core and stream stars to reconstruct this complex evolutionary history.

% WHAT RESEARCH HAS BEEN DONE WITH SAGITTARIUS
Following a host of studies involving more metal-rich stars \citep[e.g.,][]{hasselquist17,hansen18}, the Pristine Inner Galaxy Survey (PIGS) \citep{sestito24a,sestito24b,vitali24} collected medium to high-resolution spectra of the Sgr stars, including a sample of ten very metal-poor (VMP; $\ce{[Fe/H]} \le -2.0$) and two extremely metal-poor (EMP; $\ce{[Fe/H]} \le -3.0$) stars, to study the early chemical enrichment in the Sgr core. 
% The study's measurements of [Eu/Fe] show no clear trend with [Fe/H], but the sample size is small, suggesting a tentative contribution from delayed r-process sources. % like compact binary mergers.
These studies have shown the importance of examining the earliest phases of chemical evolution in Sgr and its specific nucleosynthesis event(s) to understand its assembly history.

Accordingly, we have collected a low-metallicity sample of stars in the Sgr core and stream to further investigate its early history.
We obtain a sample of 37 Sgr stars and identify ten EMP stars, 25 VMP stars, and 2 metal-poor (MP; $\ce{[Fe/H]} \le -1.0$) stars, increasing the number of EMP stars with high-resolution spectroscopy observation by a factor of five.
We find that at least half the stars in the sample, across both the core and the stream, contain \rpro\ elements whose abundances agree with the scaled solar \rpro\ pattern.
This makes Sgr another ``{\rpro}" galaxy that experienced one (or even multiple) prompt \rpro\ events at early times, prior to the formation of its stellar populations with $-3.3 < \ce{[Fe/H]} \lesssim -2$.
This highlights that some dwarf galaxies appear to be widely enriched by \rpro\ elements at early times, which so far has been limited to Reticulum\,II with possibly Tucana 3 \citep{hansen17,marshall19} and Grus II \citep{hansen20} due to their uncertain nature \citep{simon17,simon20}.

%%%%%%%%%%%%%%%%%%%%%%%%%%%%%%%%%%%%%%%%%%%%%%%%%%
\section{Target selection and observations} \label{sec:obs}

% Lambda & Beta coordindates/selection
To carry out an initial selection of the core member stars, we adopt the heliocentric Sgr coordinate system ($\Lambda_{\odot}$, $\rm{B}_{\odot}$) from \citet{majewski03} and \citet{law10}. The longitude, $\Lambda_{\odot}$, is centered on the Sgr core ($\Lambda_{\odot} = 0$\degree) and increases in the direction of the trailing Sgr debris, whereas the latitude, $\rm{B}_{\odot}$, is aligned with the debris midplane and positive toward the orbital pole ($l_{\rm GC}$, $b_{\rm GC}$) = ($273.8$\degree, $-13.5$\degree). We take the Sgr core to be located at Galactic coordinates ($l$, $b$) = ($5.6$\degree, $-14.2$\degree) and Sgr coordinates ($\Lambda_{\odot}$, $\rm{B}_{\odot}$) = (0, 0) \citep{majewski03}. The core membership is then constrained to the stars within a few degrees of the core, as seen in Table~\ref{tab:obs_param}.

% Gaia proper motion selection
We also used \Gaia\ DR3 proper motions to constrain our initial target selection and ensure a high-purity sample of Sgr member stars in its core and stream. In the core of Sgr, we first selected stars with proper motions broadly consistent with its systemic value ($-$3.6\,mas\,yr$^{-1}<\mu_\alpha<-2.7$\,mas\,yr$^{-1}$ and $-1.6$\,mas\,yr$^{-1}<\mu_\delta<-1.2$\,mas\,yr$^{-1}$), as in \citet{chf+20}.
Along its stream, we first selected stars based solely on proper motion criteria presented in \citet{ray+22}, which identified and described the proper motion trends along the Sgr tidal streams.

% photometric metallicity selection
Of our target selection, our metal-poor candidates in the Sagittarius (Sgr) dwarf galaxy were selected using metallicity-sensitive photometry from SkyMapper DR2 \citep{owb+19} and synthetic photometry derived from \Gaia\ XP spectra \citep{gaia16,gaia21,gaia23}.
For targets observed before 2023, we used photometric metallicities derived from SkyMapper DR2 in \citet{cfm+21} to select candidates in the core of Sgr.
Five low metallicity stars were observed in this manner (see Table~\ref{tab:obs_param}), but a number of original candidates ended up being contaminants with unresolved line-of-sight companions due to high median seeing at the SkyMapper site \citep{owb+19}.
With \textit{Gaia} DR3, we circumvented this issue by performing target selection using the flux-calibrated BP/RP spectra (also known as XP spectra; \citealt{carrasco21,deangeli23}) and excluding stars that had companions within 2\farcs0 based on \textit{Gaia} astrometry.
This XP-based target selection was performed by generating synthetic SkyMapper $v,g,i$ and narrow-band Ca~H\&K photometry through the Pristine survey passband (\citealt{smy+17}; analogous to \citealt{msy+24}) using the GaiaXPy toolkit\footnote{https://gaia-dpci.github.io/GaiaXPy-website/} \citep{gaia23}.
We then derived photometric metallicities following the methods in \citet{chiti20b}; our metallicities for the XP catalog have demonstrably been used to identify the lowest metallicity stars, e.g., in the LMC \citep{chiti24} and also led to the discovery of an ultra metal-poor star in the Milky Way disk \citep{mfc+24}.
% ANI confirm what was done? sameple just taken from RAY or criteria from RAY used to make your own selection
% We cross-matched our XP catalog with the final candidate sample from \citet{ray+22}.
Combining the coordinate, proper motion, and photometric metallicity selections resulted in a sample of $\sim200$ stars in the Sgr core and stream.
% $\sim20$ stars in the stream and $\sim40$ stars in the core. % this is the approximate number we actually observed

% spectroscopic observations with MIKE
We then obtained follow-up high-resolution spectroscopy of $\sim20$ stars in the stream and $\sim40$ stars in the core using the Magellan Clay $6.5$\,m Telescopes at Las Campanas Observatory using the Magellan Inamori Kyocera Echelle (MIKE) spectrograph \citep{bernstein03} in 2021 July and 2023 May and June.
All stars were observed with the 1\farcs0 slit and 2$\times$2 binning, yielding a resolving power of $R \sim 28{,}000/22{,}000$ on the blue/red arm of MIKE with wavelength coverage 3200--5000 and 4900--10000\,\AA, respectively. 
The spectra were reduced with the Carnegie Python pipeline \citep{carpy}. After shifting the spectra to rest, individual orders were normalized and stitched together into a final spectrum before analysis. Table~\ref{tab:obs_param} lists the coordinates for our observed targets, dates of their observation, exposure time, magnitudes, radial velocities, and signal-to-noise (S/N) ratios per pixel at 4500 and 6500\,{\AA}.

We derived radial velocities for these stars by cross-correlating their spectra against a template spectrum of HD122563. 
%This analysis, and visual inspection to qualitatively ascertain metallicities, was performed on-the-fly 
During data collection, to avoid observing interlopers (non-member candidates beyond a first snapshot exposure), we performed on-the-fly checks on the radial velocity and metallicity via visual inspection of the reduced exposures. Accordingly, stars with spuriously high metallicities (often resulting from the existence of unresolved companions, as discussed above) or radial velocities incongruent with membership were promptly excluded from further observations from the sample (about 20 stars).
The final sample consists of 37 stars. 

% final selection/confirmation of memership
For final confirmation of Sgr membership in the sample, we used radial velocities obtained from the total exposures and distance estimates from the Bayesian isochrone-fitting code StarHorse \citep{santiago16,queiroz18} to complete the 6D kinematics.
In the Sgr core, we broadly confirm that our stars have radial velocities in the range $100$-$180$\,\kmsec \citep{ibata94,bellazzini08}.
To confirm stream membership, we used the procedure in \citet{limberg23} that computes actions and angular momenta of the stars based on the 6D phase-space coordinates. Based on the computed actions and angular momenta of the stars, as shown in Figure~2 of \citet{limberg23}, all our stream candidates are confirmed. Since all 37 stars meet the respective criteria, our final sample consists of 22 stars with kinematics consistent with that of the core of Sgr, and 15 stars have properties in agreement with the kinematic signature of the tidal streams.

\begin{deluxetable*}{lrrrrrlrrrr} 
\tablecaption{Observational Parameters.}
\label{tab:obs_param}
\tabletypesize{\scriptsize}
\tablehead{
    \colhead{Star name} & \colhead{\Gaia\ DR3 Designation} & \colhead{R.A.} & \colhead{Decl.} & \colhead{$\Lambda_{\odot}$} & \colhead{$\rm{B}_{\odot}$} & \colhead{UT Date} & \colhead{$t_{\rm exp}$} & \colhead{$g$} & \colhead{$v_{\rm helio}$} & \colhead{S/N}
    \\
    \colhead{} & \colhead{} & \colhead{(h:m:s)} & \colhead{(d:m:s)} & \colhead{(${\rm deg}$)} & \colhead{(${\rm deg}$)} & \colhead{} & \colhead{(${\rm s}$)} & \colhead{${\rm (mag)}$} & \colhead{(${\rm km \, s^{-1}}$)} & \colhead{}% \hline
}
\startdata
\multicolumn{11}{c}{Stream} \\
\hline
{\large\noFillCircle} Sgr421   & 620763790933951488  & 09:36:00.0 & $+$17:07:21.4 & 216.9 & 11.6   & 2023 May 24 & 1320 & 14.5 & 140.1    & 32, 63 \\
{\large\noFillCircle} Sgr422   & 626434144197068160  & 09:57:30.0 & $+$18:30:17.4 & 221.7 & 9.2    & 2023 May 25 & 4163 & 16.2 & $-$61.8  & 13, 23 \\
{\large\noFillCircle} Sgr423   & 722010223233565952  & 10:21:32.3 & $+$21:46:38.8 & 226.4 & 4.6    & 2023 May 24 & 2100 & 15.0 & $-$42.8  & 32, 62 \\
{\large\noFillCircle} Sgr424   & 720678611573056384  & 10:28:03.8 & $+$20:16:15.9 & 228.3 & 5.7    & 2023 May 25 & 1500 & 15.4 & $-$41.7  & 22, 41 \\
{\large\noFillCircle} Sgr426   & 3987367858589682432 & 10:35:36.8 & $+$20:33:51.4 & 229.9 & 4.9    & 2023 May 25 & 2100 & 15.1 & $-$51.8  & 31, 62 \\
{\large\noFillCircle} Sgr428   & 3897736907644138368 & 11:49:36.0 & $+$06:15:31.9 & 252.0 & 11.8   & 2023 May 25 & 1200 & 15.9 & 193.4    & 19, 35 \\
{\large\noFillCircle} Sgr430   & 3918753557013284608 & 12:05:57.8 & $+$11:26:28.1 & 253.5 & 5.4    & 2023 May 25 & 2106 & 14.7 & $-$18.3  & 35, 82 \\
{\large\noFillCircle} Sgr431   & 3707325064494295552 & 12:36:03.1 & $+$04:22:23.4 & 263.4 & 8.3    & 2023 Jun 25 & 4794 & 16.1 & 9.8      & 18, 40 \\
{\large\noFillCircle} Sgr434   & 3708871630677912320 & 12:48:19.6 & $+$05:44:47.2 & 265.5 & 5.7    & 2023 Jun 25 & 2754 & 15.7 & $-$8.5   & 24, 43 \\
{\large\noFillCircle} Sgr438   & 3742595885684658560 & 13:17:16.3 & $+$12:51:15.1 & 268.3 & $-$4.0 & 2023 May 26 & 3990 & 15.8 & $-$103.0 & 28, 47 \\
{\large\noFillCircle} Sgr441   & 3726554388991980800 & 13:38:16.5 & $+$10:54:19.8 & 273.8 & $-$4.8 & 2023 May 26 & 5999 & 15.4 & $-$22.1  & 20, 51 \\
{\large\noFillCircle} Sgr442   & 3713742742066109568 & 13:39:13.6 & $+$04:20:10.8 & 277.3 & 0.7    & 2023 May 25 & 1800 & 15.2 & 123.2    & 29, 57 \\
{\large\noFillCircle} Sgr449   & 3724011729697503616 & 14:02:29.1 & $+$10:17:36.9 & 279.2 & $-$7.3 & 2023 May 25 & 1537 & 14.7 & $-$140.2 & 36, 66 \\
{\large\noFillCircle} Sgr452   & 3654980335256551808 & 14:34:27.4 & $+$00:57:43.5 & 290.9 & $-$3.3 & 2023 May 25 & 3386 & 16.2 & $-$7.3   & 24, 51 \\
{\large\noFillCircle} Sgr453   & 4415617329687172224 & 15:20:10.2 & $-$00:55:16.2 & 301.8 & $-$7.4 & 2023 Jun 26 & 1500 & 15.1 & $-$216.7 & 17, 41 \\
\hline
\multicolumn{11}{c}{Core} \\
\hline
{\large\fullFillCircle} Sgr459 & 4046752756534194304 & 18:35:35.3 & $-$31:04:52.3 & 356.0 & 2.9    & 2023 Jun 26 & 2400 & 15.1 & 170.3    & 23, 53 \\
{\large\fullFillCircle} Sgr460 & 4046692626988248704 & 18:37:37.0 & $-$31:12:18.7 & 356.5 & 2.9    & 2023 May 25 & 3378 & 15.9 & 154.7    & 26, 54 \\
{\large\fullFillCircle} Sgr462 & 4047106558751402112 & 18:39:44.8 & $-$30:37:02.0 & 356.8 & 2.3    & 2023 May 26 & 3600 & 15.8 & 157.4    & 29, 61 \\
{\large\fullFillCircle} Sgr463 & 4047227809930988288 & 18:43:10.7 & $-$30:13:29.2 & 357.5 & 1.7    & 2023 May 26 & 4435 & 15.0 & 161.7    & 31, 71 \\
{\large\fullFillCircle} Sgr468 & 6732306574448077824 & 18:49:32.1 & $-$34:35:45.8 & 359.6 & 5.8    & 2023 May 26 & 3964 & 15.3 & 135.3    & 30, 68 \\
{\large\fullFillCircle} Sgr2   & 6735720523692986368 & 18:49:59.0 & $-$33:49:51.5 & 359.6 & 5.0    & 2021 Jul 30 & 2673 & 15.3 & 171.9    & 24, 57 \\
{\large\fullFillCircle} Sgr471 & 6736253885603451776 & 18:50:53.7 & $-$31:33:17.7 & 359.3 & 2.7    & 2023 Jun 26 & 3000 & 15.1 & 148.0    & 28, 61 \\
{\large\fullFillCircle} Sgr476 & 6736183692971769472 & 18:52:10.3 & $-$31:54:13.2 & 359.7 & 3.0    & 2023 May 26 & 4307 & 14.9 & 166.1    & 29, 72 \\
{\large\fullFillCircle} Sgr69  & 6761301211485896192 & 18:52:48.5 & $-$29:32:23.4 & 359.4 & 0.7    & 2021 Jul 31 & 600  & 15.1 & 136.8    & 9, 26 \\ 
{\large\fullFillCircle} Sgr482 & 6760843501083241216 & 18:56:40.2 & $-$29:54:02.0 & 0.3   & 0.9    & 2023 May 25 & 2135 & 15.3 & 147.1    & 28, 58 \\
{\large\fullFillCircle} Sgr4   & 6760823641153762816 & 18:57:04.5 & $-$30:10:21.6 & 0.4   & 1.1    & 2021 Jul 30 & 1200 & 14.9 & 140.3    & 17, 47 \\
{\large\fullFillCircle} Sgr38  & 6760148545331944832 & 18:57:27.7 & $-$31:07:39.9 & 0.6   & 2.0    & 2021 Jul 31 & 3225 & 16.0 & 162.2    & 19, 45 \\
{\large\fullFillCircle} Sgr484 & 6756665051965670400 & 19:01:38.9 & $-$32:54:27.2 & 1.8   & 3.7    & 2023 May 25 & 3175 & 15.8 & 159.0    & 27, 55 \\
{\large\fullFillCircle} Sgr488 & 6755741496554745728 & 19:05:13.1 & $-$34:09:16.6 & 2.7   & 4.8    & 2023 May 26 & 2833 & 15.2 & 139.6    & 25, 57 \\
{\large\fullFillCircle} Sgr492 & 6757188660021199360 & 19:06:12.1 & $-$31:55:04.5 & 2.6   & 2.5    & 2023 May 26 & 1200 & 15.2 & 143.7    & 10, 25 \\
{\large\fullFillCircle} Sgr495 & 6757333348879294336 & 19:09:05.2 & $-$31:06:32.8 & 3.1   & 1.6    & 2023 May 25 & 1506 & 14.7 & 157.6    & 29, 67 \\
{\large\fullFillCircle} Sgr496 & 6762312242461355008 & 19:10:57.1 & $-$28:53:15.9 & 3.2   & $-$0.6 & 2023 May 25 & 2602 & 15.7 & 120.6    & 25, 55 \\
{\large\fullFillCircle} Sgr9   & 6756198442417269504 & 19:14:03.2 & $-$32:42:03.5 & 4.4   & 3.1    & 2021 Jul 30 & 2400 & 15.5 & 122.3    & 20, 50 \\
{\large\fullFillCircle} Sgr504 & 6756126841019510784 & 19:14:48.7 & $-$33:07:52.9 & 4.6   & 3.5    & 2023 Jun 26 & 1500 & 15.2 & 148.1    & 24, 47 \\
{\large\fullFillCircle} Sgr505 & 6759005362463205760 & 19:15:12.0 & $-$29:53:26.1 & 4.2   & 0.3    & 2023 Jun 26 & 3000 & 15.3 & 136.0    & 22, 47 \\
{\large\fullFillCircle} Sgr509 & 6744030735644086784 & 19:18:30.2 & $-$33:42:59.7 & 5.4   & 4.0    & 2023 May 25 & 2892 & 15.7 & 169.9    & 30, 59 \\
{\large\fullFillCircle} Sgr512 & 6744069390349796224 & 19:19:57.4 & $-$33:13:50.9 & 5.7   & 3.4    & 2023 May 25 & 2110 & 15.4 & 121.5    & 26, 55   
\enddata
\tablenotetext{}{Stars found in the Sagittarius core are denoted with black-filled circles ({\large\fullFillCircle}), and stars found the Sagittarius streams are denoted with white-filled circles ({\large\noFillCircle}). Signal-to-noise (S/N) per pixel is listed for $4500$ \AA\ and $6500$ \AA.}
% Right ascension (R.A.), declination (Decl.), and $g$ magnitude were measured by \Gaia\ using the J2000 epoch.
\end{deluxetable*}

%%%%%%%%%%%%%%%%%%%%%%%%%%%%%%%%%%%%%%%%%%%%%%%%%%
\section{Stellar parameters and chemical abundance analysis} \label{sec:stepar_abund}

%-------------------------------------------------
\subsection{Photometric stellar parameters} \label{sec:phot_stepar}

To find our first estimate of the stellar parameters (\teff, \logg, \feh, \vt) and to reaffirm that these stars are not foreground contamination, we begin by determining them spectroscopically. 
For this analysis, we used ATLAS model atmospheres from \citet{castelli03}, with $\alpha$ enhancement, as described further in Section~\ref{sec:abundances_determination}. 
We obtained the effective temperatures (\teff) by requiring no trend in the Fe\,\textsc{i} measurements as a function of excitation potential energy. Surface gravities (\logg) were derived by enforcing ionization balance between Fe\,\textsc{i} and Fe\,\textsc{ii} lines. Metallicities ([Fe/H]) were set based on the average of all available Fe\,\textsc{i} lines measured. Microturbulences (\vt) were set by requiring that Fe\,\textsc{i} have no trend with reduced equivalent width (REW). 
The process was iterated until convergence. In the end, no stars were found to have \logg\ inconsistent with being a luminous, distant red giant, further affirming their membership to Sgr.

This spectroscopic approach of deriving stellar parameters, however, typically produces cooler \teff\ and lower \logg\ compared to photometric methods for cool, metal-poor stars due to non-local thermodynamic equilibrium (NLTE) effects in Fe\,\textsc{i} \citep[e.g.,][]{frebel13,ezzeddine17}.
Using these spectroscopically derived \feh\ and \vt\ as initial guesses, we thus proceed to determine our photometric stellar parameters. 
Photometric \teff\ was derived by using \Gaia\ DR3 $G-RP$ colors \citep{gaia21} with color-[Fe/H]-\teff\ relations from \citet{mucciarelli21}.  
We retrieved photometry data from the \Gaia\ DR3 archive for the target stars and calculated the unextincted magnitudes using the extinction law provided by \Gaia \footnote{\url{https://www.cosmos.esa.int/web/gaia/edr3-extinction-law}} and the extinction map from \citep{sf11}.

\begin{figure}
    \centering
    \includegraphics[width=\linewidth]{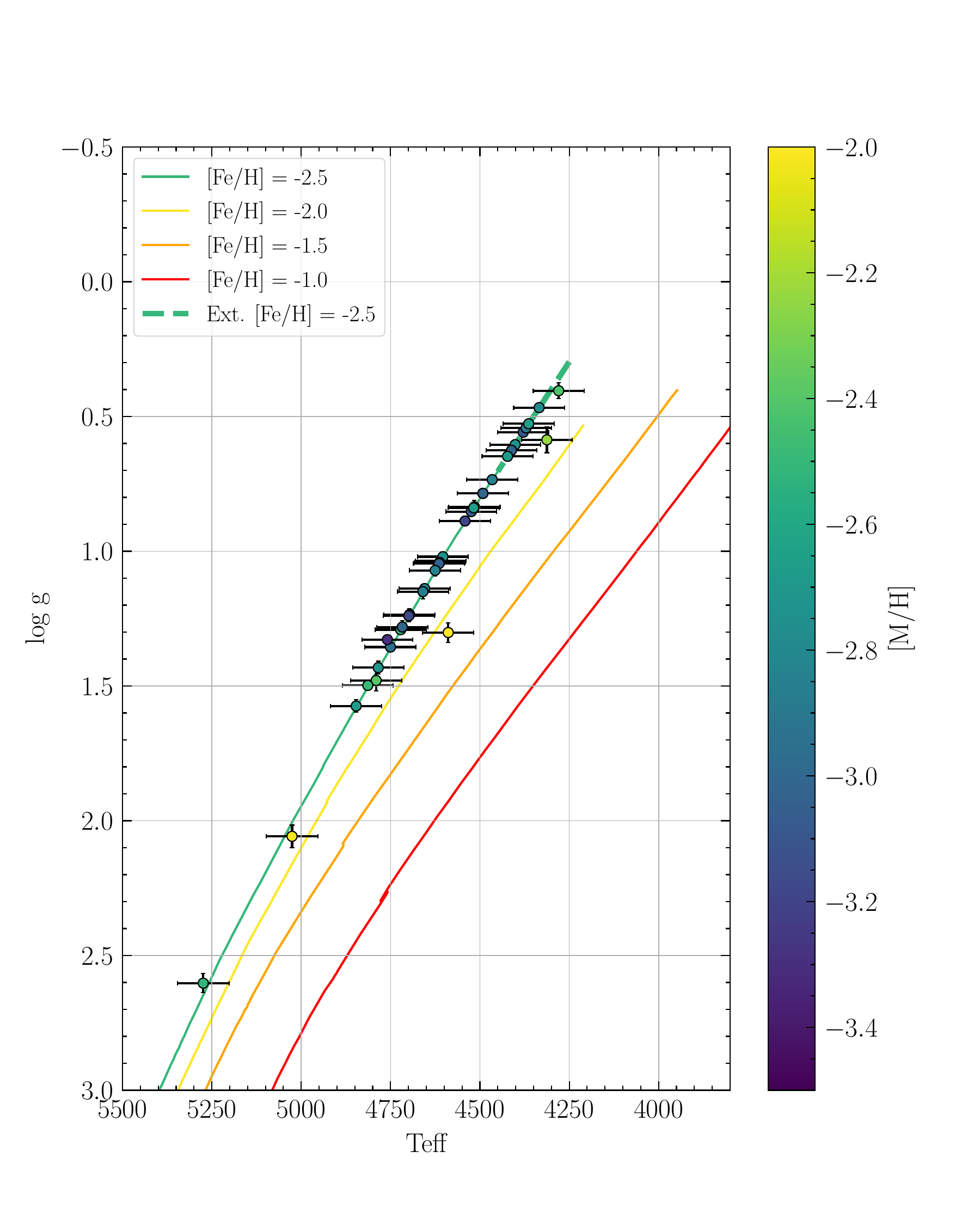}
    \caption{
        Kiel diagram for the sample with all stars shown in dots. 
        The Dartmouth isochrones are shown in solid lines.
        The dashed line represents the extrapolated isochrone.
        Both the stars and isochrones are color-coded by metallicity.
    }
    \label{fig:kiel}
\end{figure}

For our final \logg, we opted to use the isochrone fitting method rather than calculating directly from the distances.
The latter requires stars to have well-measured parallaxes (e.g., \citealp{ji20}).
Because most of the Sgr stars are at distances beyond the resolution limit for \Gaia\ parallax measurements, the astrometric distances tend to be underestimated, resulting in overestimated \logg.
We used low-metallicity Dartmouth isochrones of the red giant branch \citep{dotter08} to obtain the corresponding \logg\ for all our sample stars. For the few stars with $\rm{[Fe/H]}>-2.5$, we simply interpolate the Dartmouth isochrones in accordance to the metallicity to obtain their \logg. However, the majority of the stars have $\rm{[Fe/H]}<-2.5$, which is below the \feh\ grid limit of $\rm{[Fe/H]}=-2.5$ of the isochrones. Hence, we adopt the $\rm{[Fe/H]}=-2.5$ isochrone for these stars, knowing that the lower metallicity isochrones asymptotically converge, as shown in Figure~\ref{fig:kiel}. In addition, nine stars have photometric \teff$\lesssim4500$\,K, making them too cool to fit on the computed isochrones. We thus extrapolate the isochrone tracks by $150\,\rm{k}$ to lower \teff, assuming that the isochrone behaves qualitatively linearly towards the tip of the red giant branch.

After obtaining final values for the surface gravity and the effective temperature, we then re-determined \vt\ and \feh\ from Fe\,\textsc{i}. 
We further iterated these calculations, given that all stellar parameters depend on [Fe/H].

Statistical uncertainties on \teff\ and \logg\ were determined by propagating input uncertainties (i.e., the \Gaia\ photometry and \feh\ uncertainties) through the color-[Fe/H]-\teff\ relations and isochrone fitting, respectively. 
The \vt~statistical uncertainty was determined from the 1-$\sigma$ confidence interval in the fitted slope with respect to REW. 
Typical statistical uncertainties are $\sim72$\,K for \teff\ (dominated by uncertainties in [Fe/H]), $\sim0.02$\,dex for \logg, and $\sim0.25$\,dex for \vt.

Systematic uncertainties on \teff\ and \logg\ arise from choices in the color-[Fe/H]-\teff\ relations and isochrones.
Specifically, these relations are computed assuming certain stellar evolution and spectral energy distribution models.
The \vt\ is derived from the measured Fe lines and is thus susceptible to systematic uncertainties originating from the assumed 1D LTE stellar atmosphere models.
Consequently, we adopt systematic uncertainties of 100\,K, $0.2$\,dex, and $0.2$\,\kmsec for \teff, \logg, and \vt, respectively, which generally reflect these systematic analysis choices \citep{frebel13}.

To arrive at our total uncertainties, we add the systematic uncertainties in quadrature to the statistical uncertainties. For all stars, we find the typical total uncertainties in \teff, \logg, and \vt\ to be 
$\sim 122 \; {\rm K}$, 
$\sim 0.20 \; {\rm dex}$, and 
$\pm 0.30 \; {\rm km \, s^{-1}}$, respectively. 

The final stellar parameters and associated uncertainties are shown in Table~\ref{tab:ste_par}. 
The reported \feh\ and its associated uncertainties are computed based on the total uncertainties of \teff, \logg, and \vt.
Specifically, the uncertainties in stellar parameters introduce uncertainties in the derived element abundances. 
We propagate these uncertainties through the element abundances, including \feh, calculation following the procedure described in Section~\ref{sec:abundances_determination}.

\begin{deluxetable*}{lcccccccc}
\tablecaption{Stellar parameters of the 37 Sgr stars in this study.}
\label{tab:ste_par}
\tabletypesize{\scriptsize}
\tablewidth{\linewidth}
\tablehead{
    \colhead{Star Name} & \colhead{$T_{\rm eff}$} & \colhead{$\sigma_{T_{\rm eff}}$} & \colhead{$\log \, g$} & \colhead{$\sigma_{\log \, g}$} & \colhead{$v_{\rm t}$} & \colhead{$\sigma_{v_{\rm t}}$} & \colhead{${\rm [Fe/H]}$} & \colhead{$\sigma_{\rm [Fe/H]}$}
    \\
    \colhead{} & \colhead{(${\rm K}$)} & \colhead{(${\rm K}$)} & \colhead{(${\rm cgs}$)} & \colhead{(${\rm cgs}$)} & \colhead{(${\rm km \, s^{-1}}$)} & \colhead{(${\rm km \, s^{-1}}$)} & \colhead{(${\rm dex}$)} & \colhead{(${\rm dex}$)}}
\startdata
\multicolumn{9}{c}{Stream} \\
\hline
{\large\noFillCircle} Sgr421 & 4655 & 122 & 1.14 & 0.20 & 2.2 & 0.3 & $-$2.81 & 0.16 \\
{\large\noFillCircle} Sgr422 & 5026 & 123 & 2.06 & 0.20 & 1.9 & 0.3 & $-$1.93 & 0.19 \\
{\large\noFillCircle} Sgr423 & 4722 & 122 & 1.29 & 0.20 & 2.1 & 0.3 & $-$2.52 & 0.09 \\
{\large\noFillCircle} Sgr424 & 4814 & 122 & 1.50 & 0.20 & 1.9 & 0.3 & $-$2.43 & 0.17 \\
{\large\noFillCircle} Sgr426 & 4750 & 122 & 1.36 & 0.20 & 2.2 & 0.3 & $-$2.93 & 0.14 \\
{\large\noFillCircle} Sgr428 & 4790 & 123 & 1.48 & 0.20 & 2.2 & 0.3 & $-$2.39 & 0.18 \\
{\large\noFillCircle} Sgr430 & 4280 & 122 & 0.40 & 0.20 & 2.2 & 0.3 & $-$2.42 & 0.21 \\
{\large\noFillCircle} Sgr431 & 4659 & 123 & 1.15 & 0.20 & 2.1 & 0.4 & $-$2.82 & 0.18 \\
{\large\noFillCircle} Sgr434 & 4846 & 123 & 1.57 & 0.20 & 1.9 & 0.3 & $-$2.70 & 0.16 \\
{\large\noFillCircle} Sgr438 & 5274 & 123 & 2.60 & 0.20 & 1.4 & 0.3 & $-$2.53 & 0.16 \\
{\large\noFillCircle} Sgr441 & 4334 & 122 & 0.47 & 0.20 & 2.5 & 0.3 & $-$2.78 & 0.21 \\
{\large\noFillCircle} Sgr442 & 4697 & 122 & 1.24 & 0.20 & 2.0 & 0.3 & $-$3.21 & 0.16 \\
{\large\noFillCircle} Sgr449 & 4759 & 122 & 1.33 & 0.20 & 1.9 & 0.3 & $-$3.29 & 0.15 \\
{\large\noFillCircle} Sgr452 & 4516 & 123 & 0.84 & 0.20 & 2.4 & 0.3 & $-$3.16 & 0.17 \\
{\large\noFillCircle} Sgr453 & 4313 & 122 & 0.59 & 0.21 & 2.1 & 0.3 & $-$2.25 & 0.23 \\
\hline
\multicolumn{9}{c}{Core} \\
\hline
{\large\fullFillCircle} Sgr459 & 4363 & 122 & 0.53 & 0.20 & 2.3 & 0.3 & $-$2.66 & 0.21 \\
{\large\fullFillCircle} Sgr460 & 4717 & 123 & 1.28 & 0.20 & 2.2 & 0.3 & $-$3.00 & 0.16 \\
{\large\fullFillCircle} Sgr462 & 4625 & 122 & 1.07 & 0.20 & 2.2 & 0.3 & $-$2.83 & 0.16 \\
{\large\fullFillCircle} Sgr463 & 4423 & 122 & 0.65 & 0.20 & 2.2 & 0.3 & $-$2.69 & 0.19 \\
{\large\fullFillCircle} Sgr468 & 4491 & 122 & 0.79 & 0.20 & 2.2 & 0.3 & $-$2.57 & 0.20 \\
{\large\fullFillCircle} Sgr2   & 4609 & 122 & 1.04 & 0.20 & 2.2 & 0.3 & $-$2.66 & 0.17 \\
{\large\fullFillCircle} Sgr471 & 4542 & 122 & 0.89 & 0.20 & 2.4 & 0.3 & $-$3.14 & 0.17 \\
{\large\fullFillCircle} Sgr476 & 4371 & 122 & 0.54 & 0.20 & 1.9 & 0.3 & $-$2.88 & 0.22 \\
{\large\fullFillCircle} Sgr69  & 4401 & 122 & 0.60 & 0.20 & 2.5 & 0.3 & $-$2.68 & 0.22 \\
{\large\fullFillCircle} Sgr482 & 4604 & 122 & 1.02 & 0.20 & 2.1 & 0.3 & $-$2.74 & 0.17 \\
{\large\fullFillCircle} Sgr4   & 4379 & 122 & 0.56 & 0.20 & 2.5 & 0.4 & $-$3.10 & 0.19 \\
{\large\fullFillCircle} Sgr38  & 4784 & 123 & 1.43 & 0.20 & 2.0 & 0.3 & $-$2.72 & 0.17 \\
{\large\fullFillCircle} Sgr484 & 4697 & 123 & 1.23 & 0.20 & 2.0 & 0.3 & $-$2.94 & 0.15 \\
{\large\fullFillCircle} Sgr488 & 4492 & 122 & 0.79 & 0.20 & 2.3 & 0.3 & $-$3.02 & 0.20 \\
{\large\fullFillCircle} Sgr492 & 4466 & 122 & 0.73 & 0.20 & 2.5 & 0.6 & $-$2.81 & 0.25 \\
{\large\fullFillCircle} Sgr495 & 4412 & 122 & 0.63 & 0.20 & 2.3 & 0.3 & $-$2.96 & 0.17 \\
{\large\fullFillCircle} Sgr496 & 4613 & 122 & 1.04 & 0.20 & 1.9 & 0.3 & $-$2.95 & 0.18 \\
{\large\fullFillCircle} Sgr9   & 4589 & 122 & 1.30 & 0.20 & 2.1 & 0.3 & $-$1.85 & 0.20 \\
{\large\fullFillCircle} Sgr504 & 4517 & 122 & 0.84 & 0.20 & 2.2 & 0.3 & $-$2.67 & 0.19 \\
{\large\fullFillCircle} Sgr505 & 4524 & 122 & 0.85 & 0.20 & 2.2 & 0.3 & $-$3.06 & 0.17 \\
{\large\fullFillCircle} Sgr509 & 4699 & 122 & 1.24 & 0.20 & 2.0 & 0.3 & $-$3.18 & 0.16 \\
{\large\fullFillCircle} Sgr512 & 4614 & 122 & 1.05 & 0.20 & 2.0 & 0.3 & $-$3.03 & 0.16   
\enddata
% \tablenotetext{}{}
\end{deluxetable*}

%-------------------------------------------------
\subsection{Element abundance determination} \label{sec:abundances_determination}

We now determine the elemental abundances for all our stars. We use the 1D LTE stellar atmosphere models from \citep{castelli03} along with the radiative transfer code MOOG that includes scattering \citep{sneden73,sobeck11} and the \citet{barklem00} damping constants. 
The spectra were analyzed using \texttt{SMHR}\footnote{\url{https://github.com/andycasey/smhr}} \citep{casey14,ji20}, which provides an interface for fitting equivalent widths, interpolating the model atmospheres, and running MOOG to determine abundances from equivalent widths and spectrum synthesis. 
Equivalent widths are measured semi-automatically using a model that fits each absorption line with a Gaussian profile multiplied by a linear continuum model. 
The measurements are then manually inspected to verify each line or apply corrections. 
Abundances of synthesized lines are matched through a $\chi^2$ minimization by optimizing the abundance of a given element, the local continuum, a Gaussian smoothing kernel, and accounting for any small $v_{\rm helio}$ offsets. Each fit is again visually examined, and poor-fitting spectral regions are masked and refit. Further details are given in \citet{ji20}.

We adopted the linelist from the works of the $R$-Process Alliance with updated log $gf$ values \citep{roederer18}, generated with \code{linemake} code \citep{placco21}.
We include hyperfine splitting from \code{linemake} as well.
We assume isotopic ratios of $^{12}\rm{C}/^{13}\rm{C}=4$ given that the majority of the sample is near the tip of the red giant branch and have experienced deep mixing processes \citep[e.g.,][]{skowronski23}.
For Ba and Eu, we assume pure \rpro\ isotopes from \citet{sneden08} given the low metallicity of our sample.

\begin{figure*}
    \centering
    \includegraphics[width=0.95\linewidth]{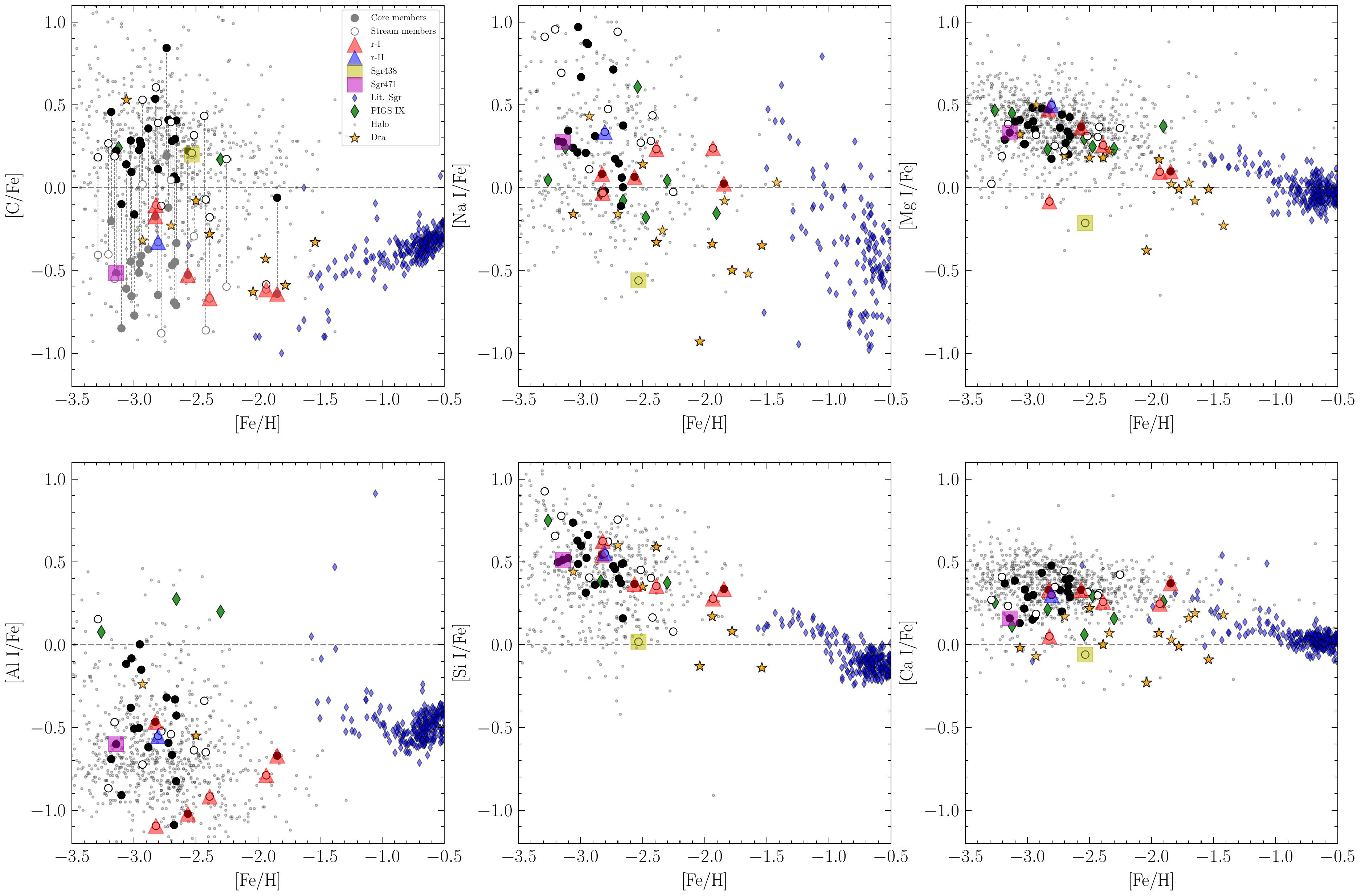}
    \caption{
    Light element abundances for the full Sgr sample.
    Stars in the Sgr streams are marked with open-face circles (\noFillCircle), whereas stars in the Sgr core are marked with black filled circles (\fullFillCircle).
    Literature values from \citet{hansen18} and APOGEE \citep{hasselquist17} are shown as transparent blue diamonds. Sgr stars from \citet{sestito24a} are shown as green diamonds (PIGS IX).
    Halo stars compiled from JINAbase \citep{abohalima18} are shown as gray dots.
    }
    \label{fig:light_elem}
\end{figure*}

\begin{figure*}
    \centering
    \includegraphics[width=0.95\linewidth]{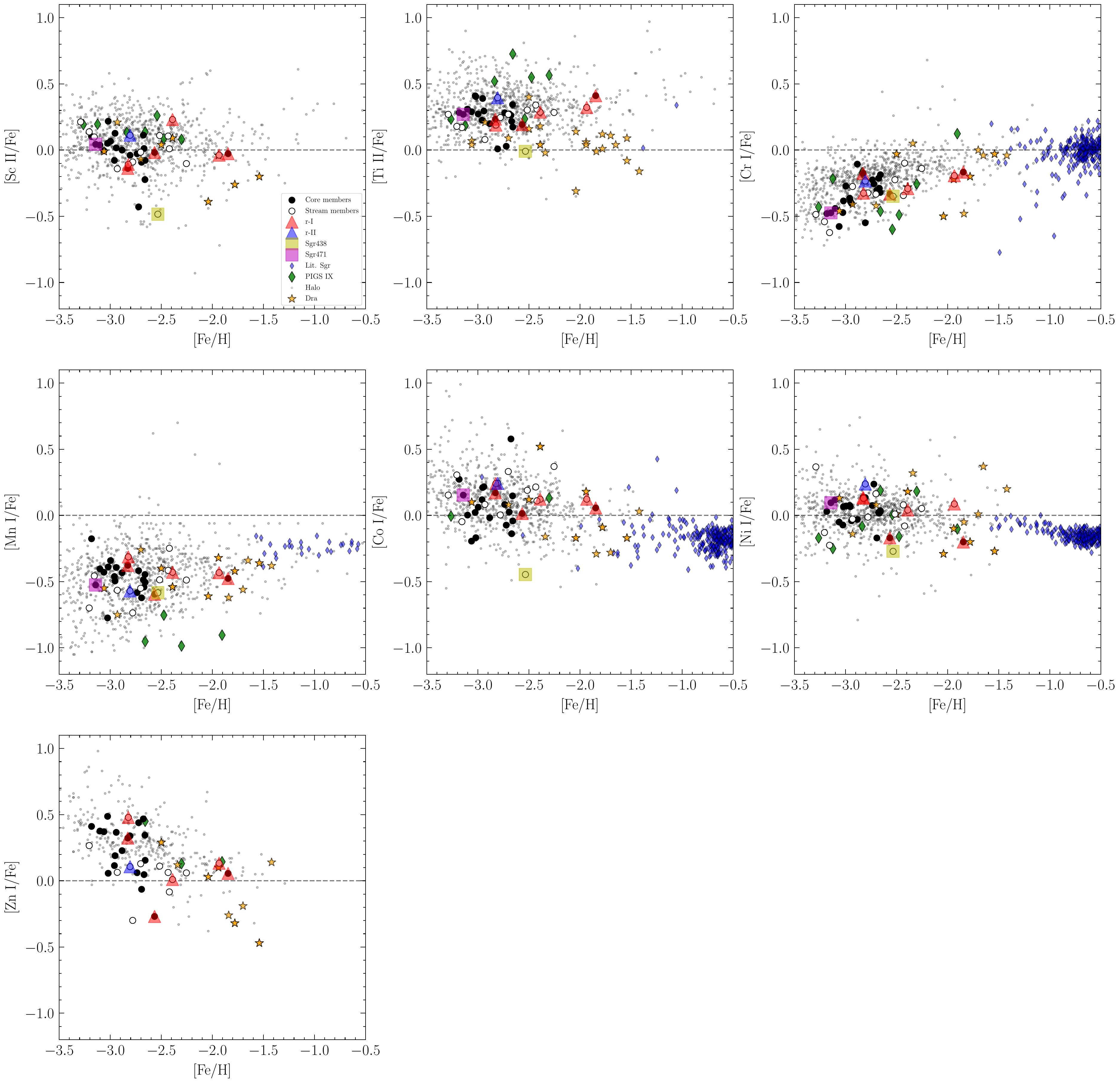}
    \caption{
    Fe-peak elements abundances for the full Sgr sample.
    Stars in the Sgr streams are marked with open-face circles (\noFillCircle), whereas stars in the Sgr core are marked with black filled circles (\fullFillCircle).
    Literature values from \citet{hansen18} and APOGEE \citep{hasselquist17} are shown as transparent blue diamonds. Sgr stars from \citet{sestito24a} are shown as green diamonds (PIGS IX).
    Halo stars compiled from JINAbase \citep{abohalima18} are shown as gray dots.
    }
    \label{fig:iron_peak_elem}
\end{figure*}

\begin{figure*}
    \centering
    \includegraphics[width=0.95\linewidth]{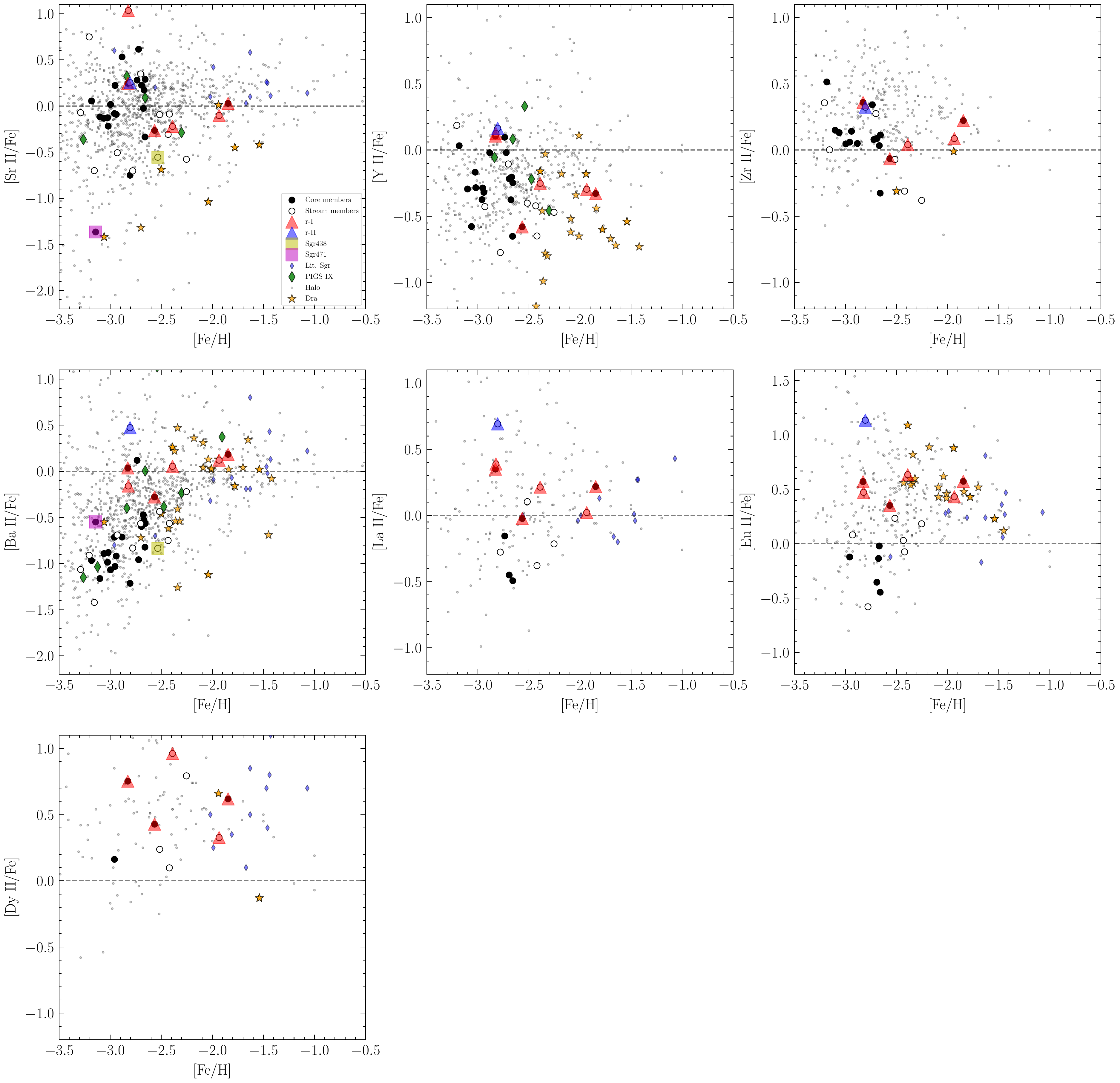}
    \caption{
    Neutron capture elements abundances for the full Sgr sample.
    Stars in the Sgr streams are marked with open-face circles (\noFillCircle), whereas stars in the Sgr core are marked with black filled circles (\fullFillCircle).
    Literature values from \citet{hansen18} and APOGEE \citep{hasselquist17} are shown as transparent blue diamonds. Sgr stars from \citet{sestito24a} are shown as green diamonds (PIGS IX).
    Halo stars compiled from JINAbase \citep{abohalima18} are shown as gray dots.
    }
    \label{fig:ncap_elem}
\end{figure*}

We measured up to 20 elements, including C, Na\,\textsc{i}, Mg\,\textsc{i}, Al\,\textsc{i}, Si\,\textsc{i}, Ca\,\textsc{i}, Sc\,\textsc{ii}, Ti\,\textsc{ii}, Cr\,\textsc{i}, Mn\,\textsc{i}, Co\,\textsc{i}, Ni\,\textsc{i}, Zn\,\textsc{i}, Sr\,\textsc{ii}, Y\,\textsc{ii}, Zr\,\textsc{ii}, Ba\,\textsc{ii}, La\,\textsc{ii}, Eu\,\textsc{ii}, and Dy\,\textsc{ii}. We provide our final derived abundances in Table~\ref{tab:abund}. The abundances of the $\alpha$-elements, as well as Na and Al, are shown in Figure~\ref{fig:light_elem}. 
The Fe-peak elements are presented in Figure~\ref{fig:iron_peak_elem}, while the neutron capture elements are shown in Figure~\ref{fig:ncap_elem}.
We note that we do not include the non-LTE correction for the following species known to be affected by departures from LTE: Na\,\textsc{i}, Al\,\textsc{i}, Cr\,\textsc{i}, and Mn\,\textsc{i}. 
This allows for a more consistent comparison of our results with those in the literature, which are largely presented without a non-LTE correction. % , and Fe\,\textsc{i}

\begin{deluxetable}{lrrrrrrr} 
\tablecaption{Chemical Abundances for the Sgr sample with the number of lines used for each element and associated total uncertainties.}
\label{tab:abund}
\tabletypesize{\scriptsize}
\tablehead{\colhead{El.} & \colhead{N} & \colhead{$\log \epsilon (\rm{X})_{\*}$} & \colhead{$\sigma$} & \colhead{${\rm [X/H]}$} & \colhead{$\sigma_{\rm [X/H]}$} & \colhead{${\rm [X/Fe]}$} & \colhead{$\sigma_{\rm [X/Fe]}$}}
\startdata
\multicolumn{8}{c}{Sgr2} \\ 
\hline
Na I  & 2   & 3.96    & 0.07      & $-$2.28 & 0.25      & 0.38    & 0.12 \\     
Mg I  & 4   & 5.25    & 0.13      & $-$2.35 & 0.16      & 0.31    & 0.08 \\     
Al I  & 2   & 3.37    & 0.22      & $-$3.09 & 0.42      & $-$0.43 & 0.29 \\     
Si I  & 2   & 5.34    & 0.21      & $-$2.17 & 0.26      & 0.49    & 0.16 \\     
Ca I  & 15  & 3.97    & 0.16      & $-$2.37 & 0.13      & 0.29    & 0.06 \\     
Sc II & 7   & 0.52    & 0.12      & $-$2.63 & 0.10      & $-$0.08 & 0.06 \\     
Ti I  & 14  & 2.30    & 0.17      & $-$2.65 & 0.20      & 0.01    & 0.07 \\     
Ti II & 29  & 2.57    & 0.15      & $-$2.38 & 0.12      & 0.17    & 0.07 \\     
V I   & 2   & 0.95    & 0.01      & $-$2.99 & 0.12      & $-$0.33 & 0.10 \\     
V II  & 2   & 1.16    & 0.10      & $-$2.77 & 0.10      & $-$0.22 & 0.11 \\     
Cr I  & 9   & 2.80    & 0.12      & $-$2.84 & 0.19      & $-$0.19 & 0.05 \\     
Cr II & 3   & 3.07    & 0.14      & $-$2.57 & 0.11      & $-$0.02 & 0.10 \\     
Mn I  & 6   & 2.27    & 0.08      & $-$3.16 & 0.11      & $-$0.51 & 0.09 \\     
Fe I  & 154 & 4.84    & 0.20      & $-$2.66 & 0.17      & 0.00    & 0.00 \\     
Fe II & 15  & 4.95    & 0.12      & $-$2.55 & 0.10      & 0.00    & 0.00 \\     
Co I  & 4   & 2.29    & 0.21      & $-$2.70 & 0.17      & $-$0.04 & 0.12 \\     
Ni I  & 4   & 3.59    & 0.21      & $-$2.63 & 0.24      & 0.03    & 0.12 \\     
Zn I  & 1   & 2.06    & 0.00      & $-$2.50 & 0.06      & 0.16    & 0.13 \\     
Sr II & 2   & 0.61    & 0.11      & $-$2.26 & 0.18      & 0.29    & 0.15 \\     
Y II  & 3   & $-$0.59 & 0.07      & $-$2.80 & 0.10      & $-$0.25 & 0.09 \\     
Zr II & 1   & 0.14    & 0.00      & $-$2.44 & 0.10      & 0.11    & 0.06 \\     
Ba II & 5   & $-$0.94 & 0.17      & $-$3.12 & 0.13      & $-$0.57 & 0.10 \\     
La II & 3   & $-$1.94 & 0.10      & $-$3.04 & 0.10      & $-$0.49 & 0.13 \\     
C-H & 2   & 5.44    & 0.01      & $-$2.99 & 0.29      & $-$0.34 & 0.19 \\     
C-N & 1   & 6.22    & 0.00      & $-$1.61 & 0.40      & 1.05    & 0.27 \\     
Eu II & 3   & $-$2.27 & \nodata & $-$2.79 & \nodata & $-$0.23 & \nodata \\
Dy II & 1   & $-$0.60 & \nodata & $-$1.70 & \nodata & 0.86    & \nodata 
\enddata
\tablecomments{%
The complete version of Table~\ref{tab:abund} is available in the online edition of the journal. 
A short version with an example star is included here to demonstrate its form and content.
}
\end{deluxetable}

% uncertainties
We estimate the uncertainties associated with our abundance measurements following the treatment detailed in \citet{ji20}, with the exception of the correlation matrix between stellar parameters (Equation~4 of \citealt{ji20}).
The correlation matrix depends on the specific stellar sample and thus should be redetermined for our Sgr sample.
We find no significant differences in the final estimated total uncertainties for most elements when not incorporating the correlation (i.e., making the matrix diagonal). 
Thus, we simply report statistical uncertainty for each element estimated from the weighted standard error in the individual line measurements for a given element. 
Systematic uncertainties are estimated by varying the model atmosphere stellar parameters based on the total uncertainties reported in Table~\ref{tab:ste_par}.
An example star (Sgr421) is shown in Table~\ref{tab:unc_example} to demonstrate the typical values for the systematic uncertainties of each element from different stellar parameter uncertainties.

\begin{deluxetable}{lrrrrr}
\tablecaption{
Systematic elemental abundance uncertainties for an example star, Sgr421 (\Gaia~DR3~620763790933951488).
}
\label{tab:unc_example}
\tabletypesize{\scriptsize}
\tablehead{
\colhead{Element} & \colhead{$\Delta_{\rm{\teff}}$} & \colhead{$\Delta_{\rm{\logg}}$} & \colhead{$\Delta_{\rm{\vt}}$} & \colhead{$\Delta_{\rm{\feh}}$} & \colhead{$\sigma_{\rm{sys.}}$} \\
\colhead{} & \colhead{$\pm 122$\,K} & \colhead{$\pm 0.2$\,dex} & \colhead{$\pm 0.3$\,\kmsec} & \colhead{$\pm 0.16$\,dex} & \colhead{}
}
\startdata
Na I & 0.15 & $-$0.05 & $-$0.16 & $-$0.04 & 0.22 \\
Mg I & 0.11 & $-$0.05 & $-$0.05 & $-$0.01 & 0.13 \\
Al I & 0.18 & $-$0.06 & $-$0.11 & $-$0.03 & 0.22 \\
Si I & 0.17 & $-$0.05 & $-$0.05 & $-$0.01 & 0.18 \\
Ca I & 0.10 & $-$0.02 & $-$0.04 & $-$0.00 & 0.11 \\
Sc II & $-$0.00 & 0.08 & $-$0.06 & 0.01 & 0.10 \\
Ti I & 0.18 & $-$0.02 & $-$0.03 & $-$0.01 & 0.18 \\
Ti II & 0.06 & 0.07 & $-$0.08 & 0.02 & 0.12 \\
V I & 0.21 & 0.02 & 0.06 & 0.05 & 0.22 \\
V II & $-$0.00 & 0.08 & 0.01 & 0.02 & 0.08 \\
Cr I & 0.16 & $-$0.02 & $-$0.02 & $-$0.01 & 0.16 \\
Cr II & $-$0.01 & 0.08 & $-$0.01 & 0.01 & 0.08 \\
Mn I & 0.09 & $-$0.04 & $-$0.01 & $-$0.01 & 0.10 \\
Fe I & 0.14 & $-$0.02 & $-$0.06 & $-$0.01 & 0.16 \\
Fe II & 0.02 & 0.08 & $-$0.03 & 0.02 & 0.09 \\
Co I & 0.14 & $-$0.03 & $-$0.03 & $-$0.01 & 0.15 \\
Ni I & 0.15 & $-$0.01 & $-$0.01 & $-$0.00 & 0.15 \\
Zn I & 0.04 & 0.03 & $-$0.01 & 0.01 & 0.06 \\
Sr II & 0.06 & 0.04 & $-$0.23 & $-$0.02 & 0.24 \\
Y II & 0.08 & 0.08 & $-$0.02 & 0.03 & 0.11 \\
Zr II & 0.04 & 0.09 & 0.01 & 0.02 & 0.10 \\
Ba II & 0.08 & 0.06 & $-$0.13 & $-$0.01 & 0.17 \\
La II & 0.07 & 0.07 & $-$0.00 & 0.03 & 0.10 \\
Eu II & 0.08 & 0.07 & 0.01 & 0.03 & 0.11 \\
Dy II & 0.02 & 0.04 & $-$0.07 & $-$0.02 & 0.08 \\
C-H & 0.28 & $-$0.07 & 0.02 & 0.06 & 0.29 \\
C-N & 0.33 & $-$0.02 & 0.06 & 0.05 & 0.34
\enddata
\end{deluxetable}

%%%%%%%%%%%%%%%%%%%%%%%%%%%%%%%%%%%%%%%%%%%%%%%%%%
\section{Discussion and Interpretation} \label{sec:discussion}

In this section, we discuss the general features of the derived chemical abundances of the Sgr sample.
We first compare our Sgr sample with the Milky Way halo, then move on to comparing with literature Sgr element abundances. We note for completeness that our sample in the Sgr stream could potentially contain halo star contaminants. 
However, without being able to further quantify the purity of our membership selection, we assume all stars as members for the purpose of our interpretation since low-metallicity interlopers from the Milky Way ought to be rare \citep{ivezic08}.

The overall distribution of light element abundances in our Sgr sample qualitatively matches that of the Milky Way halo at the same metallicity. 
This reflects extended chemical evolution in Sgr, not unlike what occurred in the early Milky Way. 
Among neutron-capture elements, we also find general agreement with halo abundance trends. 
Half of our sample show the signature of the \rpro\-process and corresponding enhancements in these elements.
This signature suggests prior enrichment, likely by a neutron star merger or another astrophysical site of the \rpro\-process.
We discuss the implications of this further in Section~\ref{sec:rpe_stars} and \ref{sec:rpro_rise}.

In Figures~\ref{fig:light_elem}-\ref{fig:ncap_elem}, we then compare our measurements with those obtained by \citet{sestito24a}, who also studied twelve very/extremely metal-poor stars in the Sgr core. 
Of the four overlapping stars between the \citet{sestito24a} and our core samples, three stars show qualitatively consistent measurements in most elements, except for Al\,\textsc{i} and Mn\,\textsc{i}.
When comparing the trends seen in both two samples, systematic differences in the measurement techniques appear to drive the differences in these two abundances.
Specifically, we adopted spectrum synthesis for measuring Al and Mn while \citet{sestito24a} used equivalent widths.
We prefer the results from spectrum synthesis due to the proximity of the two Al lines, 3944 and 3961\,\AA, to strong hydrogen lines and hyperfine structure in Mn lines.

P185053$-$313317 in their study, which is Sgr471 in our sample, appears to display a drastically higher metallicity \footnote{for unresolved reasons after checking the coordinate and potential contamination from the Moon or a second source, F. Sestito, priv. comm.}. 
While we derived $\rm{[Fe/H]}=-3.14$ for this star, \citet{sestito24a} reported $\rm{[Fe/H]}=-1.22$. 
From examining three lines, Fe\,\textsc{i} 5171.6\,\AA, Ti\,\textsc{ii} 5185\,\AA, Ti\,\textsc{ii} 5188\,\AA\ in their Figure~2 and our MIKE spectrum, we find these three lines to have a significantly weaker line depths in our spectrum. As such our abundances are commensurate with our spectrum, and their results appear to reflect their spectrum, despite quoting the same coordinates for the object. Going forward, we thus continue to use our abundances for this star. 

%==================================================

% Add discussion about Eu measurement being spurious so we are not including that
Furthermore, we investigate that five of the six stars with Eu measurements in the \citet{sestito24a} sample have detections and measurement at $\rm{[Eu/Fe]}=0.00$\,dex.
After excluding Sgr471, there are three stars left in common. 
For two stars, \citet{sestito24a} quote detections with measurements of $\rm{[Eu/Fe]}=0.00\pm0.20$\,dex based on fitting with a fixed grid of $0.1$\,dex intervals, whereas we find subsolar upper limits of $\sim-0.25$\,dex.
For internal consistency, we thus only plot Eu measurements derived in our analysis in Figure~\ref{fig:ncap_elem}.

%==================================================

In the following Sections, we further discuss individual element abundances and their interpretation. We examine our Sgr sample in the [$\alpha$/Fe] and [C/Fe] vs. \feh\ spaces as a probe of the contribution of early massive SNe in Sgr. Then we compare the abundances of the core and stream stars to assess potential differences. Finally, we report on the discovery of a population of \rpro\ enhanced stars in both the core and the stream and discuss the signatures of early \rpro\ enrichment in the Sgr progenitor. 

% ========================

\subsection{$\alpha$-element in Sgr and the star Sgr438} \label{sec:knee}

We begin by examining the $\alpha$-elements in the Sgr sample. The $\alpha$-element abundance ratios of most of our stars are enhanced at lower metallicity compared to those of the Sgr stars at higher metallicities ($\rm{[Fe/H]}\gtrsim-1.5$) from APOGEE \citep{hasselquist17,hayes20} and \citet{hansen18}.
The downturn from the higher to lower $\alpha$-abundances appears to occur in between the [Fe/H] ranges covered by the different samples but is likely due to the delayed enrichment in Fe-peak element from SNe Ia. The metallicity/location of the downturn (i.e., $\alpha$-knee) can be taken as an indicator of the stellar mass of the progenitor system. More massive systems are expected to experience more SN II contributing metals, evolving the system to a higher \feh\ faster before the SN Ia onset. 
For our sample, we see no $\alpha$-knee up to $\rm{[Fe/H]}\sim-2.0$ which suggests that it must be at $\rm{[Fe/H]}>-2.0$. This translates into a faster chemical enrichment history for Sgr compared to smaller dwarf galaxies. 
This is principally consistent with the estimate of the Sgr progenitor stellar mass of $M_*\gtrsim10^9\, M_\odot$ \citep{majewski03,niederste-ostholt10,vasiliev20}. 
Other light element ratios also show a smooth trend from lower to higher metallicity, consistent with what would be expected for the chemical evolution of a high(er) stellar mass progenitor. 
Hence, the overall trends of our Sgr stars show a typical chemical evolution history for a galaxy of its mass, especially when combined with the abundance measurements at higher metallicities.

%-------------------------------------------------

We then investigate stars that show $\alpha$-element abundances distinct from the overall sample trends.
We compute and compare the averaged $\alpha$-element measurements (Mg, Si, Ca, and Ti) to account for potential abundance variations before assessing the origins of these stars.

\begin{figure}
    \centering
    \includegraphics[width=0.95\linewidth]{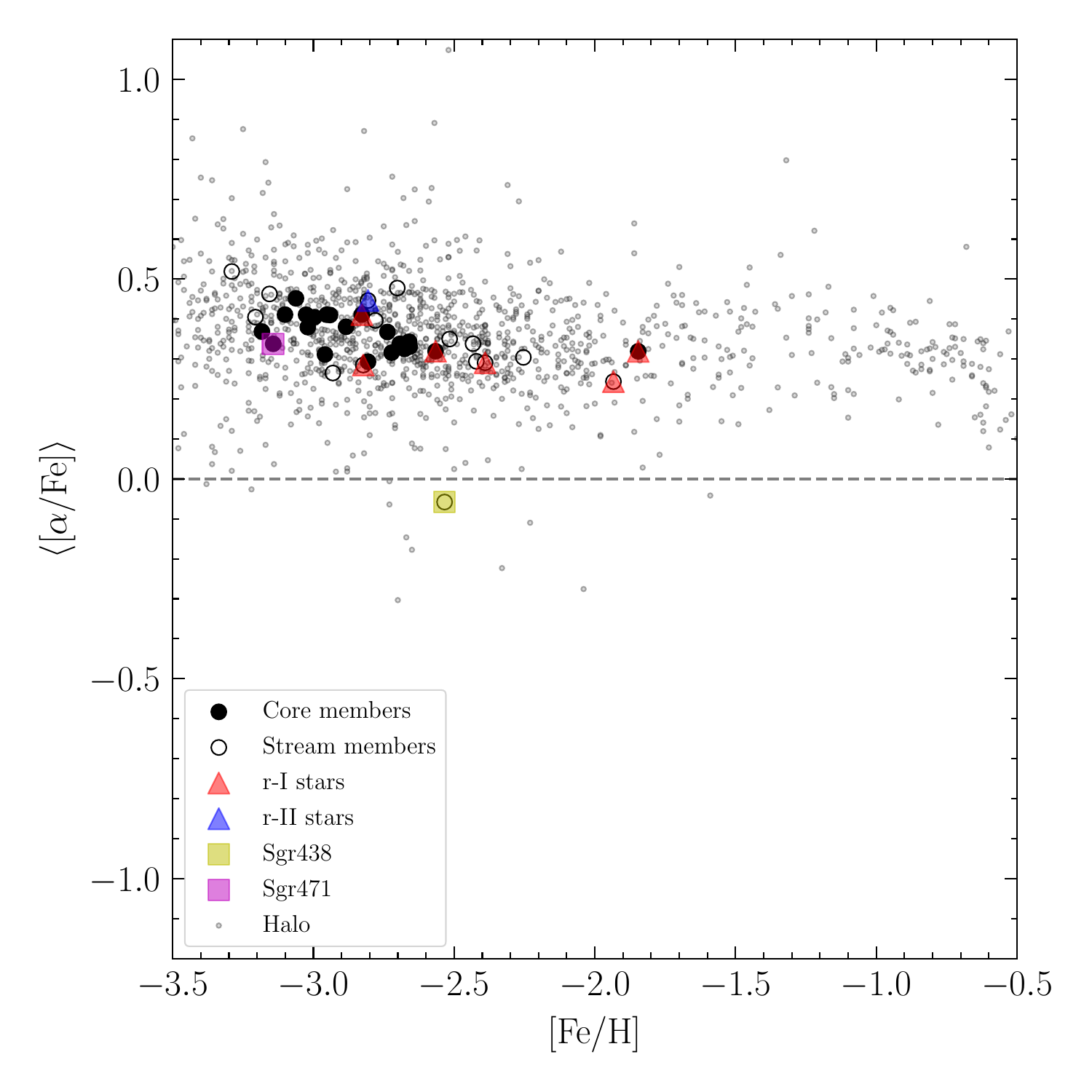} \\
    \includegraphics[width=0.95\linewidth]{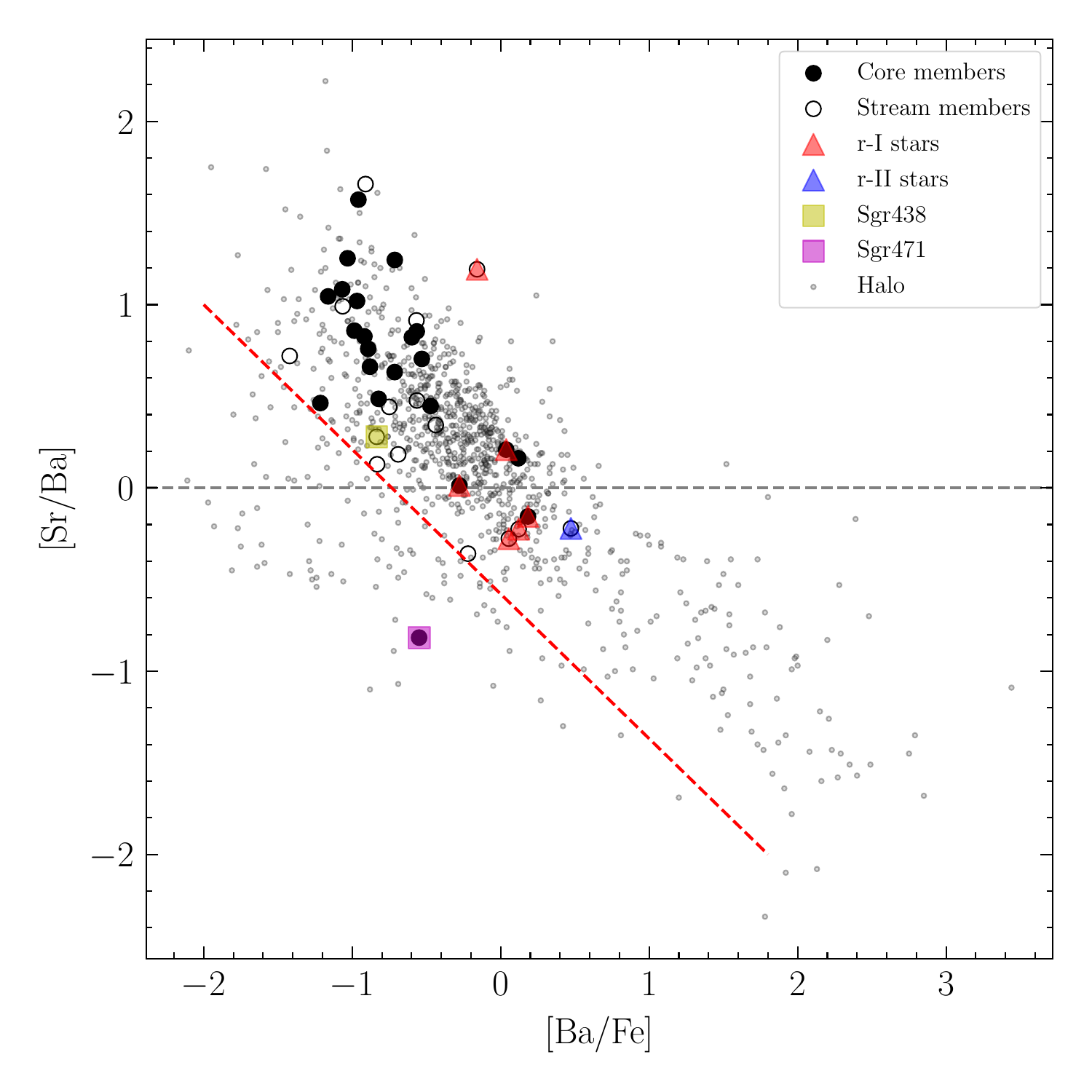}
    \caption{
    Average [$\alpha$/Fe] vs. \feh\ (top) and  [Sr/Ba] vs. [Ba/Fe] (bottom) abundances for our Sgr stars. 
    Legends are the same as Figure~\ref{fig:light_elem}.
    The top panel showcases the one star, Sgr438 with low [$\alpha$/Fe] abundances, whose origin we discuss in detail in Section~\ref{sec:knee}.
    The bottom panel shows one potential UFD star, Sgr471, with the rest of the sample distributed above the empirical cut.
    }
    \label{fig:srba_avg_alpha}
\end{figure}

Several core and stream stars have low abundances only in a single $\alpha$-element, with their average  $\alpha$-abundance broadly agreeing with the overall trend. Hence, we do not further discuss them.
But one star, Sgr438, has consistently low $\alpha$ abundances in all four elements ($\sim-0.1$), which is easily apparent in Figure~\ref{fig:light_elem} and the $\rm{[\alpha/Fe]}$ vs. \feh\ panel in Figure~\ref{fig:srba_avg_alpha}. Sgr438 has $\rm{[Fe/H]}=-2.53$ and $\rm{[Mg/Fe]}=-0.21$, $\rm{[Si/Fe]}=0.02$, $\rm{[Ca/Fe]}=-0.06$, and $\rm{[Ti/Fe]}=-0.01$ for a low combined $\rm{[\alpha/Fe]}=-0.06\pm0.07$, compared with the rest of the sample at $\rm{[\alpha/Fe]}\sim0.30$. 
Low $\alpha$ abundances are generally thought of as the results of enrichment by Type Ia SNe which contribute iron-peak elements to the gas.
Type Ia SNe contributions at relatively low metallicity suggest that our star formed in a dwarf galaxy progenitor that underwent slower chemical enrichment than what the Sgr experienced.
We thus speculate that this dwarf progenitor was less massive than the Sgr progenitor and one of its building blocks.
Likewise, the star's low Na, Sc, Co, and Sr abundances at $\rm{[X/Fe]}\sim-0.5$ show qualitative consistent behavior with measurements from the classical dwarf galaxy Draco \citep{cohen09}, as shown in Figures~\ref{fig:light_elem}-\ref{fig:ncap_elem}. Sgr438 is thus likely formed in a lower-mass dwarf progenitor with a stellar mass similar to that of Draco ($M_\star \sim 10^{5.5}\,\msun$; \citealp{odenkirchen01}) rather than in-situ. In addition, its host dwarf galaxy must have merged into Sgr before its infall into the Milky Way, given that it now belongs to the Sgr stream.

%-------------------------------------------------
\subsection{Absence of CEMP stars} \label{sec:no_cemp}

We measured the [C/Fe] abundance using the CH bandhead at 4313\,{\AA} and the feature at 4323\,{\AA}. On average, our Sgr sample of cool red giants displays $\rm{[C/Fe]}\sim-0.5$\,dex. We follow \citet{placco14} to correct the effect of evolutionary status on observed C abundances.
The correction ranges between $0.4$ and $0.7$ depending on the position of the stars along the red giant branch, i.e., \logg. The corrected [C/Fe] values are shown in the first panel of Figure~\ref{fig:light_elem}. The bulk of the sample has corrected $\rm{[C/Fe]}\sim0.2$, with a spread of $0.3\,\rm{dex}$. Interestingly, we find only one star (Sgr482) with an enhanced level of C ($\rm{[C/Fe]}=0.84$) in both the core and the stream after the correction, as shown in the first panel of Figure~\ref{fig:light_elem}. This star can now be classified as a carbon-enhanced metal-poor (CEMP) star with $\rm{[C/Fe]}>0.7$ \citep{aoki07}. The resulting fraction of CEMP star in Sgr is 1 in 37 for our sample, which is exceedingly small. This is in stark contrast to what is found among metal-poor halo stars, which generally show a large fraction of 20-$40\%$ of CEMP stars with $\rm{[Fe/H]}\lesssim-3$ \citep{placco14}. 

% mention PIGS here
We compare our [C/Fe] with those reported \citet{sestito24a} and \citet{sestito24b} and find overall good agreement for the [C/Fe] across the Sgr samples.
\citet{sestito24b} utilized medium-resolution spectroscopy for $\sim350$ metal-poor stars $\rm{[Fe/H]}\lesssim-2.5$ in the Sgr core to study the fraction of CEMP stars in the Sgr. Their corrected average [C/Fe] abundances for stars is $\langle\rm{[C/Fe]}\rangle\sim0.0\pm0.1$\,dex.
Only three or four CEMP stars were identified in the \citet{sestito24b} sample, and all are classified as potentially \spro\ enriched CEMP stars (CEMP-s), suggesting an absence of CEMP stars in the Sgr core relative to the Milky Way stellar halo population.
Similarly, comparing our C abundances with those obtained from high-resolution spectra by \citet{sestito24a}, as shown in Figure~\ref{fig:light_elem},
we find a consistent behavior across the two samples with a low to absent CEMP fraction.

% discuss the selection effect
We note that the photometric metallicity selection using the Ca\,\textsc{ii}~K feature, as is done in this study, can principally be biased against finding CEMP stars due to the presence of the CN molecular absorption feature near the Ca\,\textsc{ii}~K line and depending on the stellar parameters \citep{starkenburg17,chiti20a}.
A larger carbon abundance and lower effective temperature lead to stronger CN absorption in the spectrum. 
This can cause an artificially higher photometric metallicity measurement, making stars with strong CN features appear more metal-rich than they actually are.
Studies have shown that the most extreme CEMP stars would likely be excluded when using this technique \citep{arentsen24,chiti24}, but these selection effects likely do not completely eliminate CEMP stars (with corrected C abundances) with moderate enhancements in the \teff\ and \logg\ range of our sample, as argued in \citet{chiti24} and \citet{sestito24b}.

The differences in CEMP fraction between the Sgr and the Milky Way halo stars suggest different early chemical enrichment in the Sgr progenitor compared to the Milky Way and its progenitors.
Chemical enrichment models and SNe nucleosynthesis calculations produce enhanced [C/Fe] ratios ($\rm{[C/Fe]}>2.0$) for faint CCSNe that undergo a fall-back explosion, during which some of the Fe falls back on the nascent remnant \citep{tsujimoto03,umeda03,cooke14}. 
This results in a lower Fe yield, thus increasing the [C/Fe] ratio. 
The absence of CEMP stars in the Sgr may suggest a significant lack of faint SNe progenitors dominating the enrichment of its early gas. 
Instead, regular core-collapse SNe may rather have produced the observed carbon.
To test this idea, we use \starfit\footnote{\url{https://starfit.org/}} to fit SN nucleosynthesis yield predictions from \citet{heger10} to the average abundance patterns of both the core and stream samples.
Overall, the data was matched well, in both cases, by the yields of a relatively luminous SN with a progenitor mass of $15.8\,M_{\odot}$ and high explosion energy of $1.8\times10^{51}$\,erg. At least qualitatively, this supports that there was no significant contribution from faint CCSNe. 
A similar behavior of carbon abundances has also been observed in low metallicity stars in other dwarf galaxies in the Milky Way, such as the LMC \citep{chiti24}, Draco \citep{cohen09}, and Sculptor \citep{skuladottir24}.
%A more detailed discussion of the XXmodeling of the difference will be presented in Yelland et al. (in~prep.) %Yelland in prep. XX

%-------------------------------------------------
\subsection{Comparison of core and stream stars} \label{sec:origin}

\LCDM\ predicts that galaxies form through hierarchical assembly, with accreted stars typically residing further away from the center of the host galaxy \citep{bullock05,cooper10,cooper13}. At the same time, tidal disruption strips away stellar populations at the outskirts of the disrupted galaxy first. Thus, in the case of the Sgr, we would expect a higher fraction of stars in the Sgr stream to be accreted rather than formed in-situ compared to the stars still present in the Sgr core \citep{helmi20}. We, therefore, compare the element abundance trends between the core and stream stars to examine potential differences due to past accretion events to Sgr.

\begin{figure*}[]
    \centering
    \includegraphics[width=0.85\linewidth]{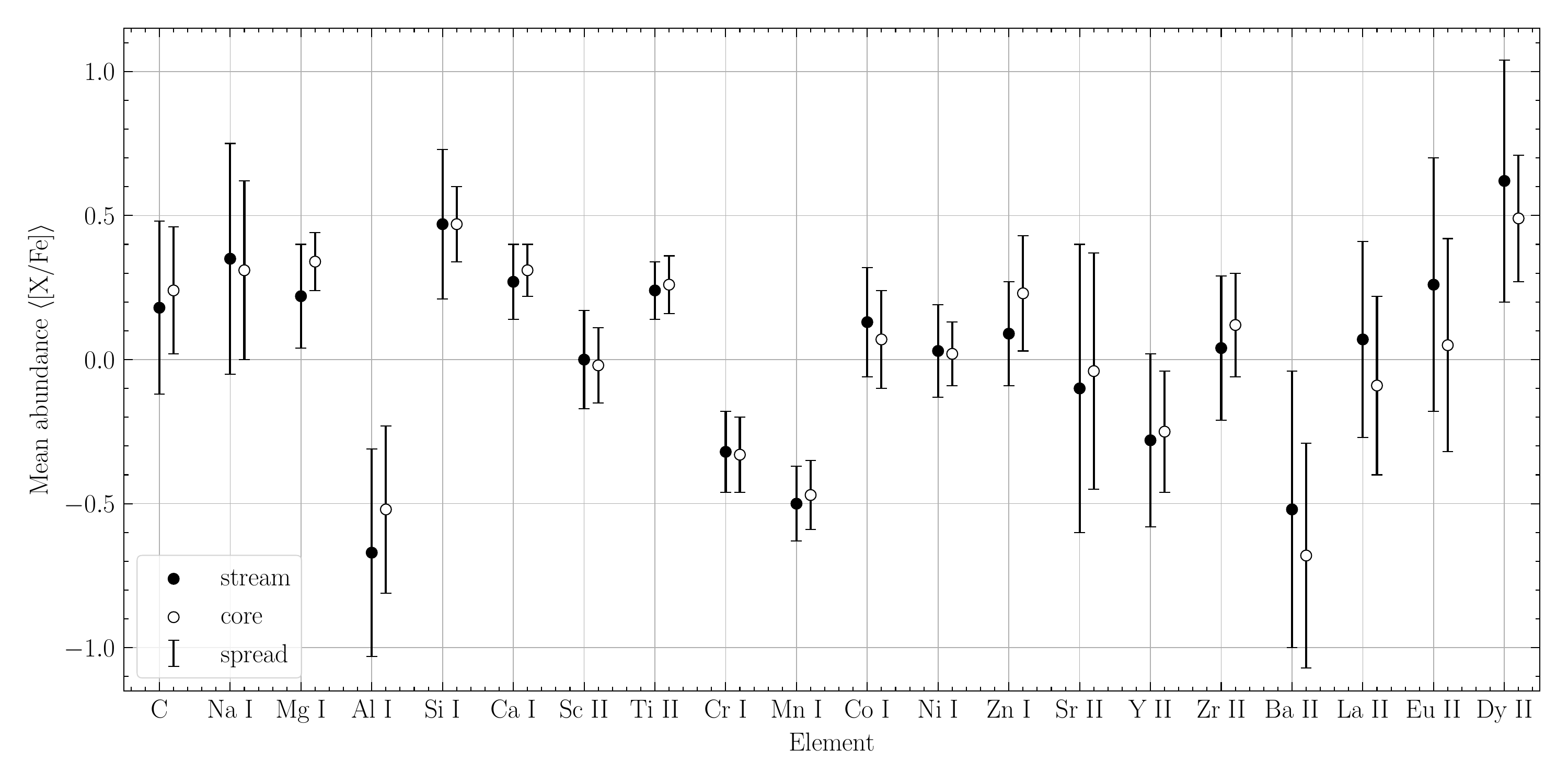}
    \caption{
    Comparison of means and spreads of various element abundances of all the stars located in the Sgr core and the stream.
    }
    \label{fig:comp_compare}
\end{figure*}

For most elements examined in this study, there are no significant abundance differences between the core and stream samples, as shown in Figure~\ref{fig:comp_compare}. 
To quantify this further, we obtained mean abundances for each sample (Table~\ref{tab:core_stream_comp}); there is no significant difference within 1$\,\sigma$ for all elements. 
These similarities suggest that the earliest progenitors that contributed to the chemical enrichment of both the early Sgr and any smaller galaxies that accreted to Sgr at early times were largely the same kind. Alternatively, similar chemical abundances could arise from a lack of major accretion events to Sgr (i.e., it evolved in isolation). The stars we observe would thus be all formed in situ, resulting in the same kind of chemical enrichment history across the core and stream populations. Either way, as a result, our metal-poor samples reflect early enrichment by SN II. 

\begin{deluxetable}{lrrrrrr}
\tablecaption{
The mean abundances and corresponding spreads for the Sgr stars in core vs. stream.
}
\label{tab:core_stream_comp}
\tabletypesize{\scriptsize}
\tablehead{
\colhead{Element} & \multicolumn{2}{c}{Mean} & \multicolumn{2}{c}{Spread} & \multicolumn{2}{c}{Number} \\
\colhead{} & \colhead{Stream} & \colhead{Core} & \colhead{Stream} & \colhead{Core} & \colhead{Stream} & \colhead{Core}
}
\startdata
% v4 abund version
% $[$C/Fe$]$ & -0.37 & -0.47 & 0.32 & 0.24 & 15 & 22 \\ % pre-correction
$[$C/Fe$]$ & 0.18 & 0.24 & 0.30 & 0.22 & 15 & 22 \\
$[$Na I/Fe$]$ & 0.35 & 0.31 & 0.40 & 0.31 & 15 & 22 \\
$[$Mg I/Fe$]$ & 0.22 & 0.34 & 0.18 & 0.10 & 15 & 22 \\
$[$Al I/Fe$]$ & $-$0.59 & $-$0.44 & 0.33 & 0.29 & 14 & 21 \\
$[$Si I/Fe$]$ & 0.45 & 0.46 & 0.26 & 0.13 & 15 & 22 \\
$[$Ca I/Fe$]$ & 0.27 & 0.31 & 0.13 & 0.09 & 15 & 22 \\
$[$Sc II/Fe$]$ & 0.01 & $-$0.01 & 0.17 & 0.13 & 15 & 22 \\
$[$Ti II/Fe$]$ & 0.24 & 0.26 & 0.10 & 0.10 & 15 & 22 \\
$[$Cr I/Fe$]$ & $-$0.32 & $-$0.33 & 0.14 & 0.13 & 15 & 22 \\
% $[$Cr II/Fe$]$ & $-$0.03 & 0.01 & 0.14 & 0.12 & 11 & 12 \\
$[$Mn I/Fe$]$ & $-$0.50 & $-$0.47 & 0.13 & 0.12 & 14 & 21 \\
$[$Co I/Fe$]$ & 0.12 & 0.06 & 0.18 & 0.15 & 15 & 21 \\
$[$Ni I/Fe$]$ & 0.03 & 0.02 & 0.16 & 0.11 & 15 & 22 \\
$[$Zn I/Fe$]$ & 0.08 & 0.22 & 0.18 & 0.20 & 12 & 20 \\
$[$Sr II/Fe$]$ & $-$0.11 & $-$0.05 & 0.49 & 0.42 & 15 & 22 \\
$[$Y II/Fe$]$ & $-$0.29 & $-$0.27 & 0.30 & 0.22 & 12 & 19 \\
$[$Zr II/Fe$]$ & 0.05 & 0.14 & 0.24 & 0.18 & 9 & 16 \\
$[$Ba II/Fe$]$ & $-$0.52 & $-$0.68 & 0.48 & 0.39 & 15 & 22 \\
$[$La II/Fe$]$ & 0.06 & $-$0.12 & 0.34 & 0.32 & 8 & 6 \\
$[$Eu II/Fe$]$ & $-$0.04 & $-$0.24 & 0.43 & 0.37 & 10 & 8 \\
$[$Dy II/Fe$]$ & 0.61 & 0.48 & 0.43 & 0.21 & 6 & 4
\enddata
\end{deluxetable}

%-------------------------------------------------
\subsection{An UFD-like star in the core: Sgr471} \label{sec:ufd_stars}

As noted, we can search for signatures of accretion from smaller dwarf galaxies onto Sgr to probe the low-mass end of the galaxy mass function. 
Stellar abundances of surviving UFD stars have been shown to display unique signatures \citep{frebel18,simon19,ji19} especially among neutron-capture elements. 
Due to their low masses, which preclude forming enough stars to sample rare enrichment events (e.g., early \rpro-process), nearly all UFDs have extremely low levels of Sr and Ba \citep[e.g.,][]{ji19}. 
They also have a characteristic lower [Sr/Ba] ratio compared with stars formed in more massive systems.
We thus examine our Sgr sample for signs of these signatures that reflect formation in UFD-mass environments. 

We find that one star, Sgr471, displays this characteristic signature in its neutron-capture elements and thus label it as likely originating from a UFD.
Specifically, Sgr471 shows a clear departure from the bulk of the Milky Way halo sample and the Sgr sample in the [Sr/Ba] vs. [Ba/Fe] space, as can be seen in the bottom panel of Figure~\ref{fig:srba_avg_alpha}.
Sgr471 has $\rm{[Fe/H]}=-2.53$, $\rm{[Sr/H]}=-4.51$, and $\rm{[Ba/H]}=-3.69$ for a low $\rm{[Sr/Ba]}=-0.82\pm0.26$.
The low [Sr/Ba] value suggests this star likely formed in a UFD environment before being accreted to Sgr. 
For completeness, we note that this star is unlikely to be a Milky Way halo foreground contaminant as it is located in the Sgr core.

Given the overall sample size of 37 stars and assuming no contaminants, we estimate a contribution of $1/37(\sim3\%)$ of the stellar mass of Sgr's low metallicity population (restricted to $\rm{[Fe/H]\lesssim-2}$) was accreted from UFDs. This estimate should be regarded as uncertain due to our small sample size and rarity of the putative accretion event.
Using the Clopper–Pearson interval \citep{clopper34}, we find the 95\% confidence interval ranges from $0.1\%$ to $14\%$ for the accreted metal-poor mass fraction of material with $\rm{[Fe/H]\lesssim-2}$. 
Nonetheless, even at $14\%$, we expect the contribution to the total stellar mass of Sgr from UFDs to be minimal as the majority of the Sgr stars have $\rm{[Fe/H]\gtrsim-2}$.
Assuming $\sim3\%$ of the Sgr stars have $\rm{[Fe/H]\lesssim-2}$, estimated from photometric metallicity distribution function from \citet{vitali22}, the UFD contribution to the total stellar mass of the Sgr can be qualitatively constrained to $\lesssim0.5\%$.

%-------------------------------------------------
\subsection{\rpro\-enhanced stars and their origin} \label{sec:rpe_stars}

\begin{figure*}
    \centering
    \includegraphics[width=0.999\linewidth]{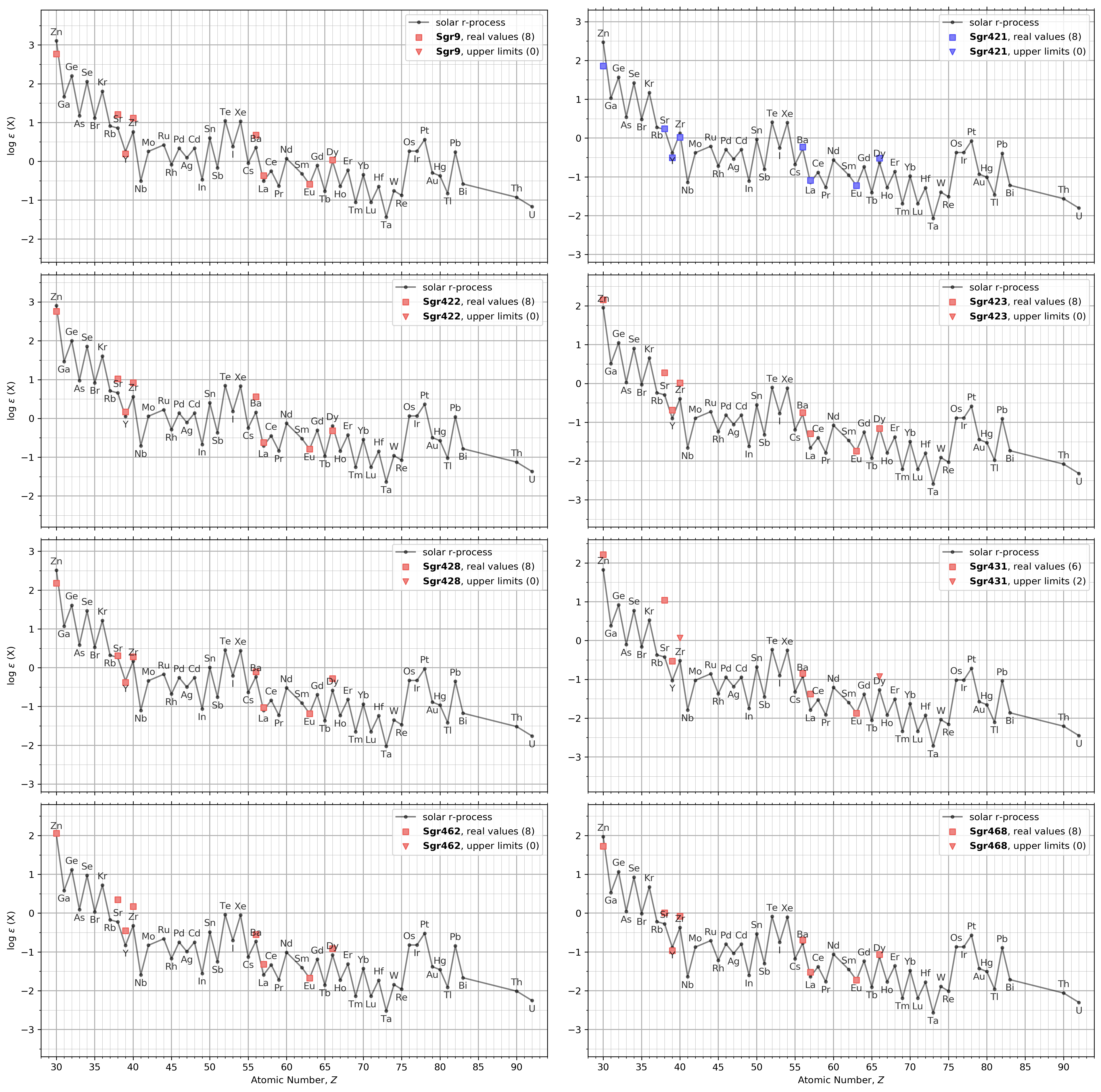}
    \caption{Heavy element abundance patterns of six r-I stars (Sgr9, Sgr428, Sgr462, Sgr431, Sgr422, Sgr468) and one r-II star (Sgr421), compared to the scaled (normalized to the stellar europium, Eu, abundance) solar \rpro\ abundance pattern (black line).}
    \label{fig:most_rproc_stars}
\end{figure*}

We now examine the overall abundance of heavy elements in stars in our sample. 
In all stars, we measure Sr, Ba; for a subset of stars, we also determine Y, Zr, La, Eu, and Dy. For these latter stars, we compare their abundance pattern with that of the scaled solar \rpro\ pattern.
This comparison reveals that Sgr does indeed contain a population of \rpro-enhanced stars.
Following the recommendation from \citet{holmbeck20}, we group \rpro-enhanced stars into two categories: r-II stars as having $\rm{[Eu/Fe]} \gtrsim 0.7$, and r-I stars as having $0.7 \gtrsim \rm{[Eu/Fe]} \gtrsim 0.3$. 

We find one r-II star (Sgr421) and six r-I stars (Sgr9, Sgr422, Sgr428, Sgr431, Sgr462, and Sgr468) in our sample. We additionally consider Sgr423 as a r-I star for the remaining discussion due to its  $\rm{[Ba/Eu]}=-0.67$ that reflects the \rpro\ ratio, and a $\rm{[Eu/Fe]}=0.24\pm0.08$ is consistent with being an r-I star within 1-$\sigma$. Five of these stars are in the Sgr stream, and three are in the Sgr core. 
The relevant element abundances are shown in Table \ref{tab:rproc_ste_par}.

\begin{deluxetable}{lrrrrl}
\tablecaption{Ba and Eu abundances of the eight \rpro-enhanced stars discussed in Section~\ref{sec:rpe_stars}.}
\label{tab:rproc_ste_par}
\tabletypesize{\scriptsize}
\tablewidth{\linewidth}
\tablehead{
    \colhead{Star Name} & \colhead{$\ce{[Fe/H]}$} & \colhead{$\ce{[Ba/Fe]}$} & \colhead{$\ce{[Eu/Fe]}$} & \colhead{$\ce{[Ba/Eu]}$} & \colhead{Classification}
    \\
    \colhead{} & \colhead{(${\rm dex}$)} & \colhead{(${\rm dex}$)} & \colhead{(${\rm dex}$)} & \colhead{(${\rm dex}$)} & \colhead{}}
\startdata
{\large\fullFillCircle} Sgr9   & $-$1.85 &    0.18 & 0.58 & $-$0.39 & r-I \\
{\large\noFillCircle} Sgr421   & $-$2.81 &    0.47 & 1.14 & $-$0.66 & r-II \\
{\large\noFillCircle} Sgr422   & $-$1.93 &    0.12 & 0.43 & $-$0.31 & r-I \\
{\large\noFillCircle} Sgr423   & $-$2.52 & $-$0.44 & 0.24 & $-$0.67 & $\sim$ r-I \\
{\large\noFillCircle} Sgr428   & $-$2.39 &    0.06 & 0.64 & $-$0.58 & r-I, $\sim$ r-II \\
{\large\noFillCircle} Sgr431   & $-$2.82 & $-$0.16 & 0.47 & $-$0.63 & r-I \\
{\large\fullFillCircle} Sgr462 & $-$2.83 &    0.04 & 0.57 & $-$0.53 & r-I \\
{\large\fullFillCircle} Sgr468 & $-$2.57 & $-$0.28 & 0.35 & $-$0.63 & r-I
\enddata
% \tablenotetext{}{}
\end{deluxetable}

These stars thus probe the origins of heavy element nucleosynthesis associated with the \rpro\ in the early Sgr. 
To investigate the \rpro\ in more detail, we compare the measured heavy element abundances of these eight stars with the scaled solar \rpro\ pattern \citep{sneden08}, as shown in Figure \ref{fig:most_rproc_stars}. 

Our confirmed \rpro\ stars cover a range of $-0.7 < \rm{[Ba/Eu]} < -0.3$.
Further, there is overall good agreement between the scaled solar \rpro\ pattern and observed abundances for elements with atomic numbers higher than Ba. 
The pure \rpro\ yields a [Ba/Eu] ratio of $-0.7$ with higher values indicating the existence of \spro, as \spro\ primarily injects Ba into the environment \citep{sneden08}.
Yet, assuming solar \rpro\ and \spro\ patterns \citep{sneden08}, the Eu contribution from \rpro\ is estimated to be $\sim150$ times that from \spro, in terms of Eu mass. 
This estimate is an over-simplification because the \spro\ pattern is somewhat metallicity dependent \citep[e.g.,][]{lugaro12}, but it nevertheless illustrates that the Eu in these eight stars is predominantly produced in \rpro\ events, confirming the operation of \rpro\ progenitor events.
% We thus expect a negligible \spro\ contribution to additional Eu in the star-forming environment, leading to a clean \rpro\ pattern.
We also examined the remaining non-rI and rII stars with available Ba and Eu values for their [Ba/Eu] to check whether their abundance patterns may reveal a \rpro contribution. 
We find over half of the remaining stars have [Ba/Eu] consistent with the scaled solar \rpro\ pattern. 

Additionally, six r-I stars (except for Sgr428) show Sr and Zr abundances higher than the scaled \rpro\ abundance. This has also been observed in other \rpro\ stars \citep{roederer10}. This excess suggests an additional contribution of CCSNe to these light neutron-capture elements, likely through a limited \rpro\ that preferentially produces Sr, Y and Zr \citep{travaglio04, honda06, wanajo11}. In addition, we find one star, Sgr431, with a Sr excess of 1.5\,dex above the scaled solar \rpro\ pattern which also suggests some early Sr producing progenitor.
Consequently, CCSNe nucleosynthesis undergoing a limited \rpro\ must have occurred prior to the first regular \rpro\ event.

The existence of eight confirmed \rpro-enhanced stars across Sgr suggests that these stars either (1) formed in the Sgr progenitor after an early \rpro\ enrichment or (2) formed in a former UFD with a clean environment before being accreted to the Sgr progenitor. 

For the first interpretation, assuming the Sgr r-I and r-II stars are formed in-situ, the Sgr must have experienced early prompt \rpro\ enrichment given their low metallicities ($\rm{[Fe/H]}\lesssim-2$). 
The early prompt \rpro\ enrichment could be neutron star mergers with short time delay and/or rare types of CCSNe \citep{siegel19,kobayashi20,yong21,cowan21,kobayashi23}.
The roughly even split between core and stream among these stars indicates no preferred sites for these early enrichment events.

For the second interpretation, we discuss what fraction of these eight stars would be expected to be accreted from UFDs. Observations of the UFDs in the local group suggest that the occurrence rate of \rpro\ events is low ($\sim1$ out of every $10$-$15$ UFDs) \citep{ji16}. 
Such a low fraction of \rpro\ enriched UFDs suggests that it is extremely unlikely that all eight stars discussed here are accretion from UFDs, i.e., $\gtrsim20\%$ ($>8$ out of 37) of the Sgr sample are accreted from UFDs. 
Specifically, we would expect to observe a factor of $10$-$15$ more stars with low neutron-capture element abundances, given the low occurrence rate of \rpro\ events in UFDs.
Given our sample size of 37 stars and the UFD-like star Sgr471 (see Section~\ref{sec:ufd_stars}) and given an estimated fraction of the accreted low metallicity population (restricted to $\rm{[Fe/H]\lesssim-2}$) from UFDs of $\sim3\%$, we estimate that plausibly one of the r-I or r-II stars could also be accreted from a UFD but not all of them. 

Taking these two potential stars accreted from UFDs, qualitatively, the accreted stellar mass fraction from UFDs to the total stellar mass remains minimal at $\lesssim0.5\%$.

%-------------------------------------------------

\subsection{Early enrichment of the \rpro\ in Sgr} \label{sec:rpro_rise}

In addition to individual stars, we examine the \rpro\ enrichment process in the full Sgr sample, following a similar procedure described in \citet{ou24c}.
Briefly speaking, the [Eu/Mg] vs. [Mg/H] space can be used to constrain the time evolution of \rpro\ sources assuming that Eu is dominantly produced by \rpro\ events, whereas Mg is dominantly produced by CCSNe events and traces the stellar formation of the galaxy.
Thus, by assuming a galaxy has a relatively smooth star formation history, we can constrain the delay time distribution and \rpro\ yield of potential \rpro\ sources (neutron star mergers and/or rare type CCSNe) by examining the stars in the [Eu/Mg] vs. [Mg/H] space.

\begin{figure}
    \centering
    \includegraphics[width=0.95\linewidth]{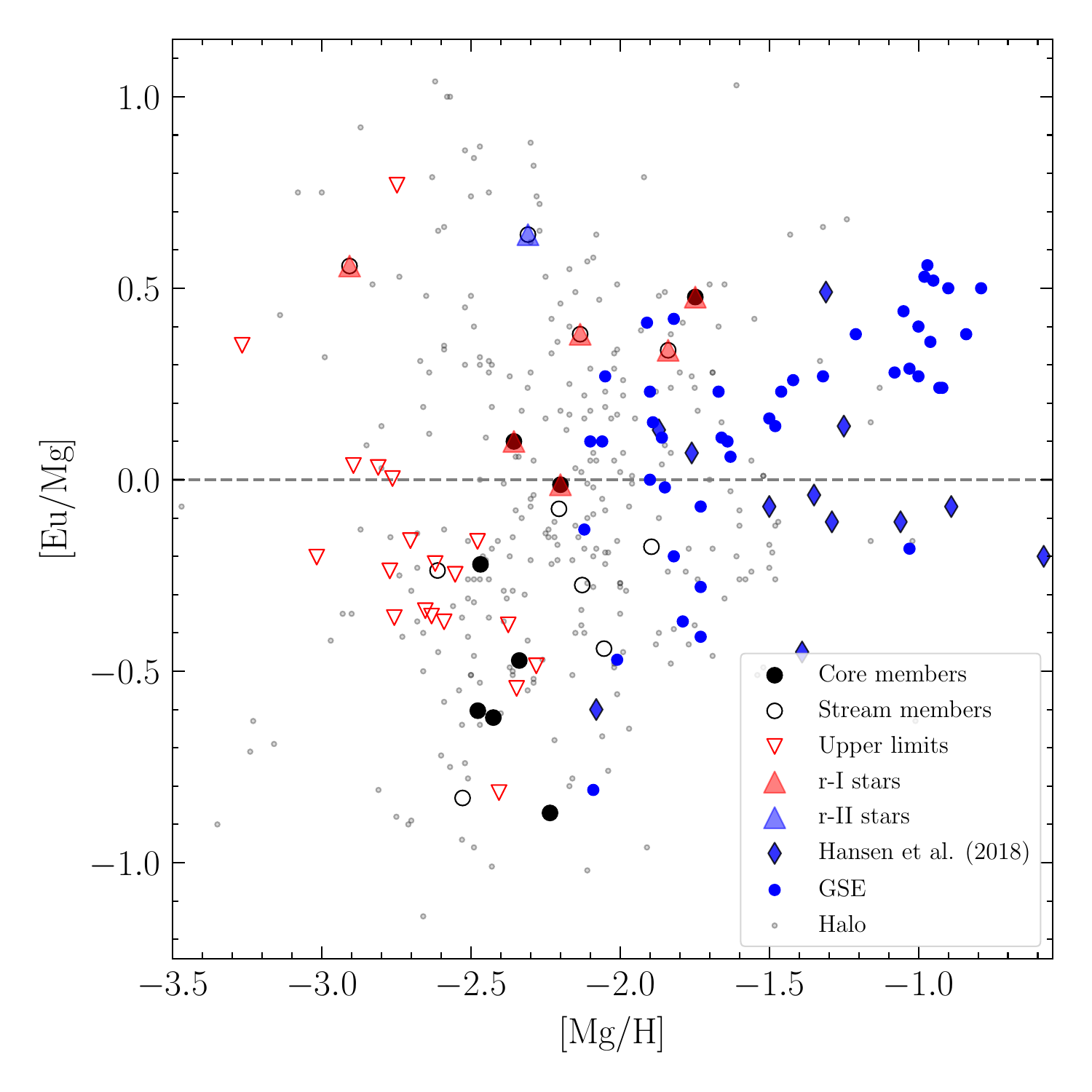}
    \caption{
    [Eu/Mg] vs. [Mg/H] for Sgr stars (black open/filled dots) in comparison with the GSE stars from \citet{ou24c} (blue dots).
    Stars in the Sgr streams are marked.
    We also include measurements from \citet{hansen18} (red diamonds), with Ca as a proxy for Mg. 
    % Sgr stars with low $\alpha$ are marked (yellow squares) as they likely have undergone different nucleosynthetic processes and may be halo contaminants.
    }
    \label{fig:eumg_mgh}
\end{figure}

From Figure~\ref{fig:eumg_mgh}, we observe a large scatter in [Eu/Mg]. 
Six of the r-I and r-II stars discussed in Section~\ref{sec:rpe_stars} have $\rm{[Eu/Mg]}\gtrsim0.0$\,dex.
The elevation in Eu relative to Mg in these r-I and r-II stars indicates the existence of early \rpro\ enrichment event(s) in the Sgr progenitor.
On the other hand, a large fraction of the sample ($\sim70\%$) has $\rm{[Eu/Mg]}\lesssim0.0$\,dex (including upper limits), resulting in an average $\rm{[Eu/Mg]}\lesssim0.0$\,dex for the full Sgr sample.

Recent observations of the LMC \citep{chiti24,oh24} and the disrupted dwarf systems show evidence of a rise in [Eu/Mg] \citep{limberg24,ou24c}, suggesting non-trivial \rpro\ contribution from delayed sources.
We thus examine our Sgr sample to search for a similar rise in Figure~\ref{fig:eumg_mgh}.

Because measurements for Eu abundances at high metallicity ([Mg/H]$\gtrsim-1.5$) are sparse and only available from \citet{hansen18}, we combine our values with the measurements from GSE stars \citep{ou24c} for comparison.
The GSE progenitor is estimated to have a similar stellar mass as the Sgr progenitor at $\sim 10^8$-$10^9$\,\msun, but the two systems had different star formation history, with the GSE forming stars for a much shorter period of time.
As a consequence, assuming the \rpro\ sources are similar in both systems, the rise should be present and span a similar range in [Eu/Mg] for Sgr and GSE.
However, in Sgr, the rise is expected to start at lower [Mg/H] and span a larger range in [Mg/H] due to its longer star formation history \citep[e.g.,][]{vitali24}.

At first glance, we find no significant evidence for a rise in our Sgr sample alone, given the relatively narrow span in [Mg/H] and large scatter in [Eu/Mg].
When combined with the GSE sample, however, the extremely metal-poor Sgr sample appears to qualitatively continue the rising trend in the GSE towards lower [Eu/Mg] and [Mg/H].
Despite the presence of the early \rpro\ enrichment events, the bulk measured [Eu/Mg] abundances for Sgr in this study are lower than the measurements for GSE stars, as well as the metal-rich sample from \citet{hansen18}.
This behavior qualitatively supports that the delayed \rpro\ sources must also be a dominant source for \rpro\ enrichment.
With more high metallicity data ($\rm{[Mg/H]}>-2.0$), we can test for the existence of the rise within the Sgr stellar population.

%%%%%%%%%%%%%%%%%%%%%%%%%%%%%%%%%%%%%%%%%%%%%%%%%%
\section{Summary} \label{sec:summary}

We investigate the chemical evolution of the Sgr dwarf galaxy by analyzing 37 stars, including 10 extremely metal-poor (EMP; \feh $\leq -3.0$), and 25 very metal-poor (VMP; $-3.0 \le$ \feh $\le -2.0$) stars. 
These stars were drawn from both the Sgr core and its tidal streams, by utilizing photometric metallicity data from SkyMapper DR2 and the \textit{Gaia} XP spectra, in addition to astrometric information from \Gaia\ DR3.
The chemical abundance patterns derived from these stars reveal several significant insights into Sgr's formation and enrichment history. 

\begin{itemize}

    \item Analysis of their $\alpha$-elements shows that most stars in the sample exhibit elevated [$\alpha$/Fe] ratios at low metallicities, consistent with rapid enrichment by core-collapse supernovae (CCSNe). 
    One star in the sample, Sgr438, exhibits significantly lower [$\alpha$/Fe] ratios, suggesting it may have originated in a low-mass classical dwarf spheroidal progenitor that merged into Sgr.

    \item A comparison of elemental abundances between the Sgr core and stream populations shows no statistically significant differences. 
    This uniformity suggests that the accreted galaxies had a similar chemical enrichment history as the Sgr progenitor, i.e., the Sgr experienced limited accretion of chemically distinct systems.

    \item The study also uncovers intriguing differences in Sgr's chemical enrichment relative to the Milky Way halo. 
    Corroborating other studies (e.g., \citealt{sestito24a, sestito24b}), we find fewer than expected CEMP stars, contrasting with the substantial fraction observed in the Milky Way halo population. 
    This indicates differences in the early stellar populations or nucleosynthesis processes that enriched the early gas in Sgr, perhaps with a lower relative contribution of faint supernovae. 

    \item One star, Sgr471, within the sample exhibits a distinct abundance pattern characteristic of stars that are found in UFDs, with notably low levels of neutron-capture element enrichment. 
    We also find eight r-I and r-II stars, and argue that $\sim1$ may plausibly be accreted from another UFD that was enriched by an \rpro\ event. 
    These discoveries highlight signatures of accretion of low-mass systems into Sgr during its assembly, consistent with hierarchical galaxy formation.

    \item Another key finding is the presence of widespread early neutron-capture element enrichment. 
    Over half of the sample shows abundance patterns consistent with the scaled solar \rpro\ pattern, with eight of them qualified as r-I or r-II stars. 
    This high fraction of stars with the \rpro\ pattern indicates that the Sgr progenitor experienced efficient \rpro\ enrichment at an early stage of its evolution. 
    Additionally, the [Eu/Mg] versus [Mg/H] abundances of the whole Sgr sample, when compared with the literature sample of more metal-rich stars from Sgr and GSE, reveals evidence for delayed \rpro\ sources contributing to the chemical evolution of Sgr.

\end{itemize}

These findings underscore the importance of studying extremely metal-poor stars to better understand the chemical enrichment history and assembly history of dwarf galaxies. 
The results reveal that Sgr experienced an enrichment history characterized by early \rpro\ events with plausible signatures of accretion from UFD-like and classical dwarf galaxy systems.

%%%%%%%%%%%%%%%%%%%%%%%%%%%%%%%%%%%%%%%%%%%%%%%%%%
\begin{acknowledgments}
X. O. thanks the LSST Discovery Alliance Data Science Fellowship Program, which is funded by LSST Discovery Alliance, NSF Cybertraining Grant \#1829740, the Brinson Foundation, and the Moore Foundation; his participation in the program has benefited this work.
X.O., A.Y., and A.F. acknowledge support from NSF-AAG grant AST-2307436. 
A.C. is supported by a Brinson Prize Fellowship at UChicago/KICP.

The national facility capability for SkyMapper has been funded through ARC LIEF grant LE130100104 from the Australian Research Council, awarded to the University of Sydney, the Australian National University, Swinburne University of Technology, the University of Queensland, the University of Western Australia, the University of Melbourne, Curtin University of Technology, Monash University and the Australian Astronomical Observatory. 
SkyMapper is owned and operated by The Australian National University's Research School of Astronomy and Astrophysics. 
The survey data were processed and provided by the SkyMapper Team at ANU. 
The SkyMapper node of the All-Sky Virtual Observatory (ASVO) is hosted at the National Computational Infrastructure (NCI). 
Development and support of the SkyMapper node of the ASVO has been funded in part by Astronomy Australia Limited (AAL) and the Australian Government through the Commonwealth's Education Investment Fund (EIF) and National Collaborative Research Infrastructure Strategy (NCRIS), particularly the National eResearch Collaboration Tools and Resources (NeCTAR) and the Australian National Data Service Projects (ANDS).

This work presents results from the European Space Agency (ESA) space mission \emph{Gaia}. \Gaia data are being processed by the \Gaia Data Processing and Analysis Consortium (DPAC). Funding for the DPAC is provided by national institutions, in particular the institutions participating in the Gaia MultiLateral Agreement (MLA). The Gaia mission website is \url{https://www.cosmos.esa.int/gaia}. The Gaia archive website is \url{https://archives.esac.esa.int/gaia}.

This research has made use of NASA's Astrophysics Data System Bibliographic Services; the arXiv pre-print server operated by Cornell University; the SIMBAD and VizieR databases hosted by the Strasbourg Astronomical Data Center.

\end{acknowledgments}

% \appendix
% % 
% This could be an appendix with sections/subsections.

%%%%%%%%%%%%%%%%%%%%%%%%%%%%%%%%%%%%%%%%%%%%%%%%%%

\bibliography{references}{}
\bibliographystyle{aasjournal}

%%%%%%%%%%%%%%%%%%%%%%%%%%%%%%%%%%%%%%%%%%%%%%%%%%

\end{document}